\documentclass[10pt,journal]{IEEEtran}
\IEEEoverridecommandlockouts
\usepackage{algorithm}
\usepackage{algpseudocode}
\usepackage{array}
\usepackage{stfloats}
\usepackage{url}
\usepackage{verbatim}
\usepackage{cite}
\usepackage{amsmath,amssymb,amsfonts}
\usepackage{graphicx}
\usepackage{textcomp}
\usepackage{siunitx} 
\usepackage{xcolor}
\usepackage{multicol}
\usepackage[nolist]{acronym, glossaries}
\usepackage{glossaries}

\usepackage{amsmath,amsfonts,bm}

\def\eqref#1{equation~\ref{#1}}

\def\1{\bm{1}}

\def\vn{{\bm{n}}}

\def\vx{{\bm{x}}}
\def\vy{{\bm{y}}}

\def\mH{{\bm{H}}}

\def\mW{{\bm{W}}}

\DeclareMathAlphabet{\mathsfit}{\encodingdefault}{\sfdefault}{m}{sl}
\SetMathAlphabet{\mathsfit}{bold}{\encodingdefault}{\sfdefault}{bx}{n}

\usepackage{todonotes}

\usepackage{tabularx,booktabs}
\usepackage{tikz}
\usepackage{pgfplots}
\pgfplotsset{compat=1.18} 
\usepgfplotslibrary{fillbetween}
\usetikzlibrary{shapes.geometric}
\usepackage{longtable}
\usepackage{color, colortbl,booktabs}
\usepgfplotslibrary{groupplots}
\usepackage{multirow}
\usepackage{subcaption}

\definecolor{lightblue}{RGB}{0, 126, 198}
\definecolor{violettblue}{cmyk}{0.9, 0.6, 0, 0}
\definecolor{lightred}{RGB}{238, 28 35}
\definecolor{lightgreen}{RGB}{140, 198, 62}
\definecolor{yellow}{RGB}{1, 221, 0}
\definecolor{orange}{RGB}{244, 111, 33}
\definecolor{lightgrey}{RGB}{128, 128, 128}

\newcommand{\sysname}{ReQuestNet }
\newcommand{\coarsenn}{CoarseNet }
\newcommand{\refinenn}{RefinementNet }

\def\BibTeX{{\rm B\kern-.05em{\sc i\kern-.025em b}\kern-.08em
    T\kern-.1667em\lower.7ex\hbox{E}\kern-.125emX}}

\begin{document}
\begin{multicols}{2}
\begin{acronym}[WSSUS]
    \acro{3GPP}{3rd Generation Partnership Project}
    \acro{5G}{fifth-generation}
    \acro{ADC}{analog to digital converter}
    \acro{AFE}{analog front end}
    \acro{AGC}{automatic gain control}
    \acro{AGV}{automated guided vehicle}
    \acro{AMP}{approximate message passing}
    \acro{API}{Application Programming Interface}
    \acro{AWGN}{additive white Gaussian noise}
	\acro{AE}{auto-encoder}
     \acro{BCE}{binary cross-entropy}
    \acro{BER}{bit error rate}
    \acro{BB}{baseband}
    \acro{bpcu}{bits per channel use}
    \acro{BP}{belief propagation}
    \acro{BPSK}{binary phase shift keying}
    \acro{BS}{base station}
    \acro{CB}{codebook}
    \acro{CE}{channel estimation}
    \acro{CL}{contrastive-loss}
    \acro{CFAR}{Constant-False-Alarm-Rate}
    \acro{CDF}{cumulative distribution function}
    \acro{CDM}{code division multiplexing}
    \acro{CFO}{carrier frequency offset}
    \acro{CoSaMP}{compressive sampling matching pursuit}
    \acro{CP}{cyclic prefix}
    \acro{CS}{compressive sensing} 
    \acro{CSI}{channel state information}
    \acro{CNN}{convolutional neural network}
	\acro{DBSCAN}{Density-Based-Spatial-Clustering-of-Applications}
    \acro{DA}{domain adaptation}
    \acro{DBA}{Distance-based Accuracy}
    \acro{DAC}{digital-analog-converter}
    \acro{DC}{direct current}
    \acro{DMRS}{demodulation reference signal}
    \acro{DE}{distance error}
    \acro{DeepL}{deep-learning}
    \acro{DoF}{degree-of-freedom}
    \acro{DFT}{discrete Fourier transformation}
    \acro{DL}{deep learning}
    \acro{DS}{delay spread}
    \acro{DGPS}{Differential Global Positioning Systems}
    \acro{DSP}{digital signal processing}
	\acro{EPS}{Constant-False-Alarm-Rate}
    \acro{ECC}{error-correcting code}
    \acro{ENoB}{effective number of bits}
    \acro{ERP}{effective radiated power}
    \acro{EVM}{error vector magnitude}
    \acro{EVD}{eigenvector decomposition}
    \acro{FB}{feedback}
    \acro{FP}{false positive}
    \acro{FN}{false negative}
    \acro{FC}{fully connected}
    \acro{FDD}{frequency division duplexing}
    \acro{FDM}{frequency division multiplexing}
    \acro{FIR}{finite impulse response}
    \acro{FFT}{fast fourier transform}
    \acro{FT}{fine tuning}
    \acro{FPGA}{field programmable gate array}
    \acro{GAN}{Generative adversarial network}
    \acro{gNB}{next generation node B}
    \acro{GPIO}{general-purpose input/output}
    \acro{GPS}{global positioning system}
    \acro{GPSDO}{GPS disciplined oscillator}
    \acro{GPU}{graphical processing unit}
    \acro{HDF}{Hierarchical Data Format}
    \acro{HDD}{hard decision decoding}
    \acro{IC}{integrated circuit}
    \acro{ICI}{inter-carrier-interference}
    \acro{ISAC}{Integrated Sensing And Communication}
    \acro{I2C}{Inter-Integrated Circuit}
    \acro{ICSP}{in-circuit serial programming}
    \acro{IF}{intermediate frequency}
    \acro{i.i.d.}{independent and identically distributed}
    \acro{IIR}{infinite impulse response}
    \acro{IMU}{inertial measurement unit}
    \acro{IoT}{Internet of Things}
    \acro{IPS}{indoor positioning system}
    \acro{IR}{infrared}
    \acro{JSDM}{Joint Spatial Division and Multiplexing}
    \acro{LiDAR}{Light Detection And Ranging}
    \acro{LLR}{log-likelihood ratio}
    \acro{LP}{leakage precoder}
    \acro{LMMSE}{Linear Minimum Mean Square Error}
    \acro{LO}{local oscillator}
    \acro{LoS}{line of sight}
    \acro{LiDaR}{Light Detection and Ranging}
    \acro{LM}{likelihood module}
    \acro{LS}{least squares}
    \acro{LSTM}{long-term short-term memory}
    \acro{LTE}{Long Term Evolution}
    \acro{LTI}{linear time invariant}
    \acro{LTV}{linear time variant}
    \acro{MAP}{maximum a posteriori}
    \acro{ML}{maximum likelihood}
    \acro{MMSE}{minimum mean squared error}
    \acro{mmWave}{millimetre Wave}
    \acro{MUSIC}{Multiple Signal Classification}
    \acro{NN}{Neural Network}
    \acro{MLP}{multilayer perceptron}
    \acro{NNI}{Neural Network Intelligence}
    \acro{NLoS}{non-line of sight}
    \acro{KNN}{k-nearest neighbors}
    \acro{OCC}{orthogonal cover codes}
    \acro{OOD}{out of distribution}
    \acro{OFDM}{orthogonal frequency division multiplex}
    \acro{PRB}{physical resource block}
    \acro{PRG}{physical resource block group}
    \acro{RADAR}{Radio Detection And Ranging}
    \acro{RGB}{Red-Green-Blue}
    \acro{ReLU}{rectified linear unit}
    \acro{RB}{resource block}
    \acro{RE}{resource element}
    \acro{RIM}{Recurrent inference machine}
    \acro{RF}{radio frequency}
    \acro{RMS-DS}{Root Mean Square - Delay Spread}
    \acro{RNN}{recurrent neuronal network}
    \acro{RSSI}{received signal strength indicator}
    \acro{R-ZF}{regularized zero-forcing}
    \acro{SDD}{soft decision decoding}
    \acro{SDR}{software defined radio}
    \acro{SE}{spectral efficiency}
    \acro{SFO}{sampling frequency offset}
    \acro{STO}{sampling time offset}
    \acro{SLAM}{Simultaneous Localization and Mapping}
    \acro{SGD}{stochastic gradient descent}
    \acro{SISO}{single input single output}
    \acro{SINR}{signal-to-interference-and-noise-ratio}
    \acro{SIR}{signal-to-interference-ratio}
    \acro{SLNR}{signal-to-leakage-and-noise ratio}
    \acro{SNR}{signal-to-noise-ratio}
    \acro{SP}{subspace}
    \acro{SQR}{signal-to-quantization-noise-ratio}
    \acro{SQNR}{signal-to-quantization-noise-ratio}
    \acro{SVD}{singular value decomposition}
    \acro{SU}{single-user}
    \acro{TDD}{time division duplexing}
    \acro{TRIPS}{time-reversal IPS}
    \acro{TRP}{transmission and reception point}
    \acro{TP}{true positive}
    \acro{TN}{true negativ}
    \acro{UE}{user equipment}
    \acro{UL}{uplink}
    \acro{ULA}{uniform line array}
    \acro{URLLC}{ultra-reliable low-latency communication}
    \acro{US}{uncorrelated scattering}
    \acro{USRP}{universal software radio peripheral}
    \acro{UWB}{ultra-wideband}
    \acro{WiFi}{Wireless Fidelity}
    \acro{WSS}{wide sense stationary}
    \acro{WSSUS}{wide sense stationary uncorrelated scattering}
	\acro{YOLO}{You-Only-Look-Once}
    \acro{ZF}{zero forcing}
    \acro{MSE}{Mean Squared Error}
    \acro{NMSE}{Normalized MSE}
    \acro{TDL}{tapped delay line}
    \acro{CDL}{clustered delay line}
    \acro{LOS}{Line of Sight}
    \acro{NLOS}{Non-LOS}
    \acro{SCS}{Subcarrier spacing}
\end{acronym}
\end{multicols}

\title{
ReQuestNet: A Foundational Learning model for Channel Estimation
}


\author{
    \IEEEauthorblockN{
        Kumar Pratik\IEEEauthorrefmark{1}, Pouriya Sadeghi\IEEEauthorrefmark{2}, Gabriele Cesa\IEEEauthorrefmark{1}, \\
        Sanaz Barghi\IEEEauthorrefmark{2}, Joseph B. Soriaga\IEEEauthorrefmark{2}, Yuanning Yu\IEEEauthorrefmark{2},
        Supratik Bhattacharjee\IEEEauthorrefmark{2}\IEEEauthorrefmark{3}, Arash Behboodi\IEEEauthorrefmark{1}
    }\\
    \IEEEauthorblockA{
        \IEEEauthorrefmark{1}Qualcomm AI Research\thanks{
            Qualcomm AI Research is an initiative of Qualcomm Technologies, Inc. \\
            Corresponding author: Kumar Pratik (\texttt{kpratik@qti.qualcomm.com})\\
            ©2025 Qualcomm Technologies, Inc. and/or its affiliated companies. All Rights Reserved.
        }, Amsterdam,
        \IEEEauthorrefmark{2}Qualcomm Technologies Inc., USA.
        \thanks{\IEEEauthorrefmark{3}Work completed while affiliated with Qualcomm Technologies Inc., USA.}
    }
}

\maketitle
\vspace{-3mm}

\begin{abstract}
In this paper, we present a novel neural architecture for channel estimation (CE) in 5G and beyond, the Recurrent Equivariant UERS Estimation Network (ReQuestNet). It incorporates several practical considerations in wireless communication systems, such as ability to handle variable number of \ac{RB}, dynamic number of transmit layers, physical resource block groups (PRGs) bundling size (BS), demodulation reference signal (DMRS) patterns with a single unified model, thereby, drastically simplifying the CE pipeline. Besides it addresses several limitations of the legacy linear MMSE solutions, for example, by being independent of other reference signals and particularly by jointly processing MIMO layers and differently precoded channels with unknown precoding at the receiver. \sysname comprises of two sub-units, \coarsenn followed by RefinementNet. \coarsenn performs per PRG, per transmit-receive (Tx-Rx) stream channel estimation, while \refinenn refines the \coarsenn channel estimate by incorporating correlations across differently precoded PRGs, and correlation across multiple input multiple output (MIMO) channel spatial dimensions (cross-MIMO). 
Simulation results demonstrate that \sysname significantly outperforms genie \ac{MMSE} CE across a wide range of channel conditions, delay-Doppler profiles, achieving up to 10dB gain at high SNRs. Notably, \sysname generalizes effectively to unseen channel profiles, efficiently exploiting inter-PRG and cross-MIMO correlations under dynamic PRG BS and varying transmit layer allocations.
\end{abstract}

\section{Introduction}\label{sec:introduction}
The advent of 5G NR and the anticipated evolution toward sixth-generation (6G) networks have ushered in an era of unprecedented connectivity, data throughput, and system complexity. These developments necessitate advanced techniques for low-power, compute-efficient, and reliable wireless communication. \ac{CE} is a fundamental building block of modern wireless communication systems, and high quality \ac{CE} is essential for achieving the target throughput required by many applications. Orthogonal Frequency Division Multiplexing (OFDM), a foundational modulation scheme in 5G NR, creates parallel communication channels across a large time-frequency grid. To acquire \ac{CSI}, the pilot signals known as \ac{DMRS} is used, whose time-frequency positions and values are known a priori to both transmitter and receiver. The CE task involves estimating CSI at non-DMRS resource elements (REs) from noisy DMRS observations. The standard allows for a variety of DMRS patterns \cite{channels2020modulation}, tailored to different use cases and deployment scenarios. The CE needs to work across various system configurations and channel conditions, and to run efficiently for all hundreds of channels within the time-frequency resource grid. Techniques such as variable configuration for DMRS and leveraging other reference signals are additional overheads for the 5G stack, and will get more cumbersome as the channel dimensionality increases. A better channel estimation algorithm can help to trade off some of the existing overheads, for example by decreasing DMRS density, while keeping the CE quality at the desired level.

Acquiring the channel statistics,  adapting the underlying CE algorithm accordingly, and maximize the processing gain by joint estimation across the whole resource grid are among the challenges of traditional methods.  
Linear \ac{MMSE} based solutions are among the best methods for the channel estimation, given their optimality for jointly Gaussian estimation problems.
However, MMSE methods rely on precise knowledge of channel and noise statistics—information that needs to be acquired beforehand. It is important to note that when MMSE methods are done jointly over a grid of thousands of resouruces, they incur high computational cost due to the need for second-order statistics and matrix inversion. To address this, several works \cite{neumann2018learning} have proposed approximations that reduce complexity by estimating, rather than computing, channel statistics. Besides, as we will show, these methods cannot address the joint estimation of differently precoded channels and deal with broken \ac{OCC}. MMSE methods are example of parameter-based CE frameworks that typically follow a two-stage process: first, estimating channel-related parameters (e.g., Doppler spread, delay profile, noise variance) using reference signals such as PT-RS and CSI-RS; second, adapting filter parameters based on these estimates. While effective in controlled settings, these methods struggle to adapt to the dynamic and heterogeneous propagation environments of 5G NR. Challenges include varying DMRS configurations (e.g., configuration type, number of OFDM symbols), dynamic PRG bundling (e.g., bundle size, precoding strategy), and multiple \ac{SCS}. High user mobility, dense base station deployments, and rich multipath propagation further exacerbate the problem. MIMO systems introduce additional complexity increasing both the dimensionality of the estimation task and the spatial interference.

Our goal is to provide a design for a unified channel estimation algorithm to address these challenges, that is, it does not depend on other reference signals and aims at maximizing the processing gain by joint CE across the whole grid. The models aims at harvesting gains where the best linear estimation methods fail, particularly, for joint channel estimation across differently precoded resource groups (PRGs) and joint processing of MIMO layers. The goal is to build a realistic channel estimator and substantiate the gain with respect to the best linear estimators.  
Thereby, we hope to open the discussion around overhead reduction for channel acquisition. Our design should be foundational: it should be able to be applied or adapted to variety of design ideas for pilot patterns, channel conditions and MIMO configurations.  In this paper, we propose a novel neural architecture for parameter-free CE in 5G NR and beyond. We design a neural network architecture, given the promises of deep learning in high-dimensional, non-linear, and dynamic inference tasks, and with their intrinsic differentiability. We combine 
various ideas from deep learning based inverse problems, permutation equivariant networks, transformers, recurrent and convolutional architecture to design a model that follows best practices in the machine learning field and matches the desiderata of channel estimation.  We show that with proper data curation and training curriculum, the network outperforms genie linear MMSE CE for in and \ac{OOD} tasks. Although the complexity of the model, currently around a few million parameter, can be optimized further for on device deployment, we would like to highlight that the network can also be used for assistance in exploring more innovative and efficient ways of designing reference signals and resource grids, as well as classical channel estimation methods.

\subsection{Related Works}
A wide range of neural network architectures have been explored for CE in 5G-OFDM systems, including Multi-Layer Perceptrons (MLPs) \cite{mlbased5gCE, dlaided5GCE, dlforCE}, Convolutional Neural Networks (CNNs) \cite{drlforCE, dlbasedCE, Cammerer2023-wj}, Recurrent Neural Networks (RNNs), hybrid models \cite{cnn_rnn_CE, pinns_ce}, and attention-based frameworks \cite{attentionCE}, and many others \cite{Fischer2022-za,Wiesmayr2024-jf}.

\textbf{MLP-based approaches:} MLPs have been used to model CE either end-to-end—directly mapping received signals to transmitted symbols \cite{dlforCE}—or as auxiliary modules to enhance traditional estimators such as LS \cite{dlaided5GCE}.

\textbf{CNN-based approaches:} CNNs are effective in capturing spatial and time-frequency correlations in the channel response. In \cite{neumann2018learning}, a CNN was trained to approximate MMSE estimation by learning its parameters. ChannelNet \cite{dlbasedCE} employed a super-resolution CNN to interpolate full CSI from pilot positions. ReEsNet \cite{drlforCE} introduced a deep residual CNN for per $T_x$–$R_x$ CE.

\textbf{RNN-based approaches:} RNNs, particularly LSTM and BiLSTM variants, have been applied to model temporal dependencies in time-varying channels \cite{rnn_ce}, making them suitable for high-mobility scenarios.

\textbf{Hybrid models:} Combining CNNs and RNNs enables joint modeling of spatial and temporal correlations. For instance, \cite{cnn_rnn_CE} proposed a hybrid architecture where CNNs extract spatial features and RNNs capture temporal dynamics, achieving state-of-the-art performance.

\textbf{Attention-based models:} Attention mechanisms have been introduced to dynamically focus on relevant input features. In \cite{attentionCE}, a transformer-based encoder was paired with a residual decoder to enhance CE accuracy, outperforming conventional neural architectures.

\textbf{Physics-Informed Neural Networks (PINNs):} PINNs integrate domain knowledge into the learning process by embedding physical constraints. The framework in \cite{pinns_ce} demonstrated improved generalization and robustness by combining data-driven learning with physical modeling, offering a principled approach to CE.

\subsection{Contributions and Paper Outline}

This paper makes the following key contributions:

\begin{itemize}
    \item We propose ReQuestNet, a foundational DL architecture for parameter-free channel estimation in 5G NR systems, capable of operating across diverse configurations including varying \ac{DMRS} patterns, PRG bundling strategies, dynamic transmit layers, and \acp{SCS}.
    
    \item We introduce a novel recurrent estimation framework that jointly processes MIMO spatial streams and differently precoded PRGs. This capability is formally linked to the multi-reference alignment problem, providing a principled explanation for the model’s ability to align and integrate structurally misaligned channel observations.
    
    \item We design a modular and permutation-equivariant architecture that incorporates learned iterative inference, attention-based inter-PRG and cross-MIMO modeling, and a hybrid coarse-to-fine estimation pipeline, enabling robust generalization across channel conditions and deployment scenarios.
    
    \item We conduct extensive experiments across modularity, standardization, and generalization settings, demonstrating that a single trained instance of \sysname consistently outperforms genie-adided classical MMSE estimator, and generalizes effectively to \ac{OOD} channel models such as \ac{CDL}.
\end{itemize}

The remainder of the paper is organized as follows. Section~\ref{sec:background} introduces the system model and relevant background on 5G NR CE. Sections~\ref{sec:guiding_principles} and \ref{sec:requestnet-dl} present the guiding principles and architectural details of ReQuestNet. Section~\ref{sec:experiments} describes the training methodology and experimental setup, followed by a comprehensive evaluation across multiple scenarios. Finally, Section~\ref{sec:conclusions} concludes the paper and outlines future research directions.
\section{Background}\label{sec:background}

\subsection{Notation}\label{sec:notation}
The following notations are used throughout the paper. Scalars are denoted by lowercase letters (e.g., $x$), vectors by bold lowercase letters (e.g., $\mathbf{x}$), and matrices by bold uppercase letters (e.g., $\mathbf{H}$). The identity matrix of size $n$ is denoted by $\mathbf{I}_n$. The superscript $t$ on a variable (e.g., $\hat{\mathbf{H}}^t$) denotes the $t$-th decoding iteration, while subscripts $(j,k)$ refer to the receive-transmit antenna pair. The notation $\hat{\mathbf{H}}$ denotes an estimated channel matrix, while $\mathbf{H}$ denotes the ground truth. The operator $\dagger$ denotes the Hermitian transpose. Unless otherwise specified, all channels are assumed to be complex-valued and unit-powered. The terms spatial stream, and $T_x$–$R_x$ link are used interchangeably to refer to a single MIMO transmission path.

\subsection{5G Numerology}\label{sec:numerology}
We begin by outlining how time, frequency, and antenna dimensions are structured in 5G NR. The standard employs OFDM as its primary modulation scheme, with a flexible frame structure designed to support diverse service requirements. A 5G NR frame spans $10$ ms, divided into $10$ subframes of $1$ ms each. Each subframe contains one or more slots, depending on the numerology, i.e., \ac{SCS}.

\begin{figure}
    \centering
    \includegraphics[width=0.7\linewidth]{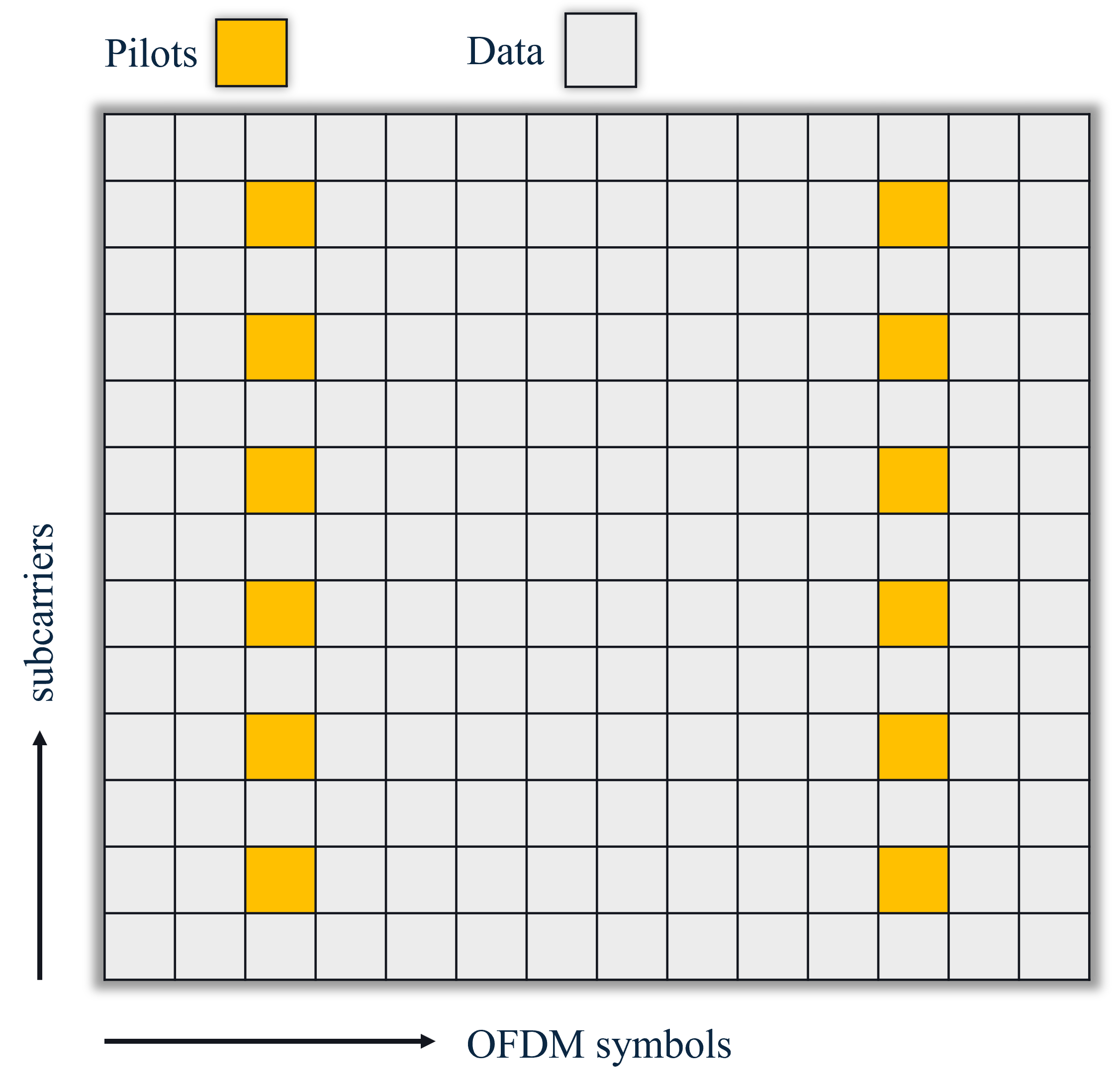}
    \caption{\textit{An example 5G frame.} Schematic of a 5G NR Physical Resource Block (PRB), with subcarriers along the frequency axis (y-axis) and OFDM symbols along the time axis (x-axis). Yellow REs denote DMRS allocations; remaining REs correspond to data transmission.}
    \label{fig:5g_frame_example}
\end{figure}

The resource grid is a time-frequency lattice where each slot is subdivided into OFDM symbols (time domain) and subcarriers (frequency domain). The number of slots per subframe increases with \ac{SCS}, while the number of OFDM symbols per slot depends on the cyclic prefix (CP): $14$ symbols for normal CP and $12$ for extended CP. The resource grid is structured as follows:

\begin{itemize}
    \item \textbf{Subcarrier spacing:} \ac{SCS} values include $15$, $30$, $60$, $120$, and $240$ kHz depending on the numerology.
    \item \textbf{OFDM symbols:} Each slot contains $14$ symbols (normal CP) or $12$ symbols (extended CP).
    \item \textbf{\ac{RB}:} Each \ac{RB} comprises $12$ subcarriers over one slot as illustrated in Fig.~\ref{fig:5g_frame_example}.
\end{itemize}

Fixing these parameters yields a time-frequency grid of resource elements (REs), each used for MIMO-OFDM transmission. A group of \acp{RB} is allocated for communication. For further details, we refer our readers to \cite{Chen2021-ej}.

\subsection{Narrowband precoding}\label{sec:narrowband_precoding}
Precoding is a fundamental operation in MIMO systems, aligning the transmit symbol vector $\mathbf{x}$ with the eigenstructure of the channel matrix $\mathbf{H}$. This is typically achieved by multiplying $\mathbf{x}$ with a precoding matrix derived from the singular value decomposition of $\mathbf{H}$. Ideally, each \ac{RE} would use a distinct precoding matrix to account for channel variations across time and frequency. However, this approach is impractical due to two main challenges: (i) the overhead of acquiring fine-grained downlink CSI at the base station, and (ii) the increased complexity of CE when precoders vary across \acp{RE} and are unknown at the receiver. To address this, 5G NR supports two precoding strategies. In narrowband precoding, a shared precoding matrix is applied across a small group of \acp{RB} (typically $2$ or $4$), optimizing transmission over a narrow frequency band. In contrast, wideband precoding applies a single matrix across the entire bandwidth.

The concept of \ac{PRG} bundling governs the selection of precoding granularity. A \ac{PRG} groups multiple \acp{RB} into a bundle, with the bundle size (BS) determining the frequency span over which a single precoder is applied. Supported BS values in 5G NR include $2$, $4$, and wideband. PRG bundling enhances flexibility by enabling distinct precoding matrices for different PRGs, thereby improving spectral efficiency under varying channel conditions.

\subsection{Reference signals}\label{sec:reference_signals}
5G NR defines several reference signals to facilitate channel estimation, synchronization, and demodulation, including CSI-RS, TRS, PTRS, and \ac{DMRS}. Among these, \ac{DMRS} is specifically designed for estimating the precoded downlink channel. It consists of predefined pilot symbols transmitted on a subset of \acp{RE}, known to both transmitter and receiver. The receiver uses these pilots to interpolate the channel across the full set of allocated \acp{RB}. Multiple \ac{DMRS} configurations—such as Type I/II and single/dual-symbol variants—determine the time-frequency density of pilots, balancing robustness to channel dynamics (e.g., Doppler shift) with pilot overhead \cite{Chen2021-ej}. An example pattern is shown in Fig.~\ref{fig:5g_frame_example}.

To support MIMO transmission, 5G NR employs DMRS ports, which represent distinct transmit pilot streams. These are separated using \ac{OCC} or cyclic shifts, enabling the receiver to isolate per-layer channels by exploiting spatial orthogonality. This design assumes that adjacent \acp{RE} experience similar channel conditions, allowing effective suppression of inter-layer interference and accurate per-layer CE.

\subsection{Channel Estimation Problem }\label{sec:problem_setup}
Consider a 5G NR OFDM system with $N_{tr}$ transmit and $N_r$ receive antennas forming a MIMO configuration. The input-output relationship for pilot transmissions is given by:
\begin{equation}
\mathbf{y} = \mathbf{H} \mathbf{P} \mathbf{x} + \mathbf{n}, \label{eq:system_model}
\end{equation}
where $\mathbf{y} \in \mathbb{C}^{N_r}$ is the received signal, $\mathbf{H} \in \mathbb{C}^{N_r \times N_{tr}}$ is the MIMO channel matrix, $\mathbf{x} \in \mathbb{C}^{N_{tr}}$ is the transmitted symbol vector, and $\mathbf{n} \sim \mathcal{CN}(0,\sigma^2\mathbf{I}_{N_r})$ is complex Gaussian noise. The objective of \ac{CE} is to estimate $\mathbf{H}$ at non-pilot locations using the received signal $\mathbf{y}$ and known pilot symbols $\mathbf{x}$.

While CE has been extensively studied in wireless communications, 5G NR introduces new challenges. Estimating the channel across the full \acp{RB} requires understanding how the channel evolves between pilot symbols in both time and frequency, which depends on the underlying delay-Doppler profile. This necessitates adaptive estimation strategies that can track and respond to varying channel statistics. Although 5G NR provides reference signals such as \ac{DMRS} to aid in this process, the estimator must still infer and adapt to these statistics in real time.

Another challenge arises from spatial correlation and the breakdown of orthogonality between DMRS ports, particularly in high delay spread environments. The assumption that adjacent \acp{RE} experience similar channels becomes less valid, degrading the effectiveness of \ac{OCC} and complicating per-layer estimation. Additionally, narrowband precoding introduces further complexity. In 5G NR, the same precoding matrix is applied across a group of \acp{RB} defined by the PRG BS. Larger PRG BS enables joint processing across more \acp{RB}, potentially improving CE robustness through increased processing gain. However, this comes at the cost of reduced frequency resolution for precoder selection. Conversely, smaller PRG BS improves frequency resolution but limits joint processing due to the lack of known correlation across PRGs, as the \ac{UE} does not know the exact precoders used by the \ac{gNB}.

This trade-off between CE quality and precoder optimization resolution also limits the effectiveness of traditional LS and MMSE estimators, whose interpolation windows are constrained by the PRG BS. Narrow PRG bundling restricts the estimator’s ability to exploit correlations across frequency, often resulting in a significant performance gap compared to wideband CE.

These challenges motivate several key questions:
\begin{itemize}
    \item Can a single, parameter-free model perform robust \ac{CE} across diverse channel conditions and NR configurations relying solely on \ac{DMRS}?
    \item Can CE performance be improved under inter-layer interference caused by MIMO correlation and high delay spread?
    \item Can wideband-like processing gain be achieved in narrowband precoding scenarios without knowledge of the applied precoders?
\end{itemize}

In this work, we answer these questions affirmatively by proposing a unified, data-driven CE model for 5G NR that generalizes across configurations and channel conditions.

\subsection{CE as Multi-Reference Alignment}\label{subsec:ce_mra}

The CE problem under unknown narrowband precoding can be interpreted as an instance of the multi-reference alignment (MRA) problem. Consider two received signals:
\begin{align}
   \vy^{(1)} & = \mH^{(1)} \mW^{(1)} \vx_{dmrs}^{(1)} + \vn^{(1)} \\
   \vy^{(2)} & = \mH^{(2)} \mW^{(2)} \vx_{dmrs}^{(2)} + \vn^{(2)},
\end{align}
where $\mW^{(1)}$ and $\mW^{(2)}$ are unknown, independently chosen square precoding matrices. Since $\mH^{(1)} \mW^{(1)}$ and $\mH^{(2)} \mW^{(2)}$ are uncorrelated, traditional MMSE estimators cannot jointly process them. However, we can rewrite the second equation as:
\[
   \vy^{(2)} = \mH^{(2)} \mW^{(1)} (\mW^{(1)})^{-1} \mW^{(2)} \vx_{dmrs}^{(2)} + \vn^{(2)},
\]
where $(\mW^{(1)})^{-1} \mW^{(2)}$ represents the unknown transformation between the two precoding matrices. This formulation reveals that joint estimation involves recovering $\mH^{(1)} \mW^{(1)}$, $\mH^{(2)} \mW^{(1)}$, and the relative transformation between precoders—an instance of the MRA problem, where observations are misaligned by unknown transformations.

This is analogous to the heterogeneous Cryo-EM problem \cite{Singer2020-af}, where different molecular structures are observed under unknown rotations. In our case, $\mH^{(1)}$ and $\mH^{(2)}$ represent different channel realizations, and $\mW^{(i)}$ are unknown linear transformations (precoders). While the CE problem is less noisy than Cryo-EM, it is complicated by the heterogeneity of the channels and the fact that precoders may not be square or invertible. Our proposed model addresses these challenges by learning to align and jointly estimate channels across PRGs without requiring explicit knowledge of the precoders.

\section{Guiding principles and high level overview of the ReQuestNet architecture}\label{sec:guiding_principles}
We begin by outlining the core principles that guide the design of ReQuestNet, grounded in both the mathematical formulation of the CE problem and the practical constraints of wireless communication systems. We then provide a high-level overview of the architecture, illustrating how these principles are embedded into the network's structure.

\subsection{Guiding principles}
\sysname is designed to address key challenges in 5G NR channel estimation that are difficult to resolve using classical methods such as linear \ac{MMSE}. Below, we describe three foundational principles that inform the architecture.

\subsubsection{Learned Inverse Problem Solvers}
We model pilot-based MIMO transmission as a linear system: $\mathbf{y} = \mathbf{H}\mathbf{x} + \mathbf{n}$, where $\mathbf{x}$ is the known pilot, $\mathbf{H}$ is the unknown precoded channel, and $\mathbf{y}$ is the received signal. Channel estimation can be posed as a \ac{MAP} inference problem:
\begin{equation}
\max_{\mathbf{H}} \log \text{p}\left(\mathbf{H}|\mathbf{y,x}\right) \propto \max_{\mathbf{H}} \left(\log \text{p}(\mathbf{y}|\mathbf{H,x}) + \log \text{p}_{\theta}(\mathbf{H})\right),
\label{eq:main_MAP_estimation}
\end{equation}
where $\text{p}(\mathbf{y}|\mathbf{H,x})$ is the likelihood and $\text{p}_\theta(\mathbf{H})$ is a prior over the channel, potentially parameterized by a neural network. This formulation generalizes the linear \ac{MMSE} estimator under Gaussian assumptions. Recent advances in learned inverse problem solvers demonstrate that such MAP formulations can be effectively addressed using deep learning. One approach is to learn the prior $\text{p}_\theta(\mathbf{H})$ using generative models such as GANs, normalizing flows, or diffusion models \cite{Bora2017-oa,Whang2021-wp,Lei2019-lj,Zirvi2024-ju,Daras2024-iy,Chung2022-wj}. Another approach is to learn the optimization process itself using iterative neural networks \cite{Gregor2010-ij,Behboodi2022-ls,Schnoor2023-mx,Zhang2018-yx,Liu2019-ci,Putzky2017-sq,Lonning2018-wr}. We adopt the latter strategy by modeling the iterative MAP update:
\begin{equation}
\mathbf{H}^{t+1} = \mathbf{H}^t + \gamma_t \nabla \left(\log \text{p}\left(\mathbf{y}|\mathbf{H}^t\right) + \log \text{p}_{\theta}\left(\mathbf{H}^t\right)\right),
\label{eq:map_update}
\end{equation}
where $\gamma_t$ is the step size at iteration $t$. The \ac{RIM} framework \cite{Putzky2017-sq} learns to perform such updates without requiring an explicit prior. Instead, it uses a neural update operator:
\begin{equation}
\mathbf{H}^{t+1} = \mathbf{H}^{t} + h_{\mu}^{\mathbf{H}}\left(\nabla_{\mathbf{y|H}^t}, \mathbf{H}^t\right),
\label{eq:map_update_2}
\end{equation}
where $h_{\mu}^{\mathbf{H}}(\cdot)$ is a learnable function and $\nabla_{\mathbf{y|H}^t} = \nabla \log \text{p}(\mathbf{y}|\mathbf{H}^t)$. The classical update in Eq.~\eqref{eq:map_update} is recovered when:
\begin{equation}
h_{\mu}^{\mathbf{H}}\left(\nabla_{\mathbf{y|H}^t}, \mathbf{H}^t\right) = \gamma_t \left(\nabla_{\mathbf{y|H}^t} + \nabla \log \text{p}_{\theta}(\mathbf{H}^t)\right).
\label{eq:map_exp}
\end{equation}

This principle forms the foundation of ReQuestNet’s refinement mechanism, enabling it to learn iterative updates that incorporate both data fidelity and learned structural priors.

\subsubsection{Permutation Equivariance}
A neural \ac{CE} model can be viewed as a function $\mathbf{f}_{\theta}(\mathbf{y}, \mathbf{x}) \rightarrow \hat{\mathbf{H}}$, which estimates the channel matrix $\mathbf{H}$ given the received signal $\mathbf{y}$ and the transmitted pilot $\mathbf{x}$. In MIMO systems, the ordering of transmit-receive spatial streams is arbitrary—permuting the columns of $\mathbf{H}$ and the corresponding entries of $\mathbf{x}$ does not alter the received signal $\mathbf{y}$. Therefore, the channel estimator should be permutation equivariant with respect to the ordering of spatial streams.

\begin{figure}[!ht]
    \centering
    \includegraphics[width=0.8\linewidth]{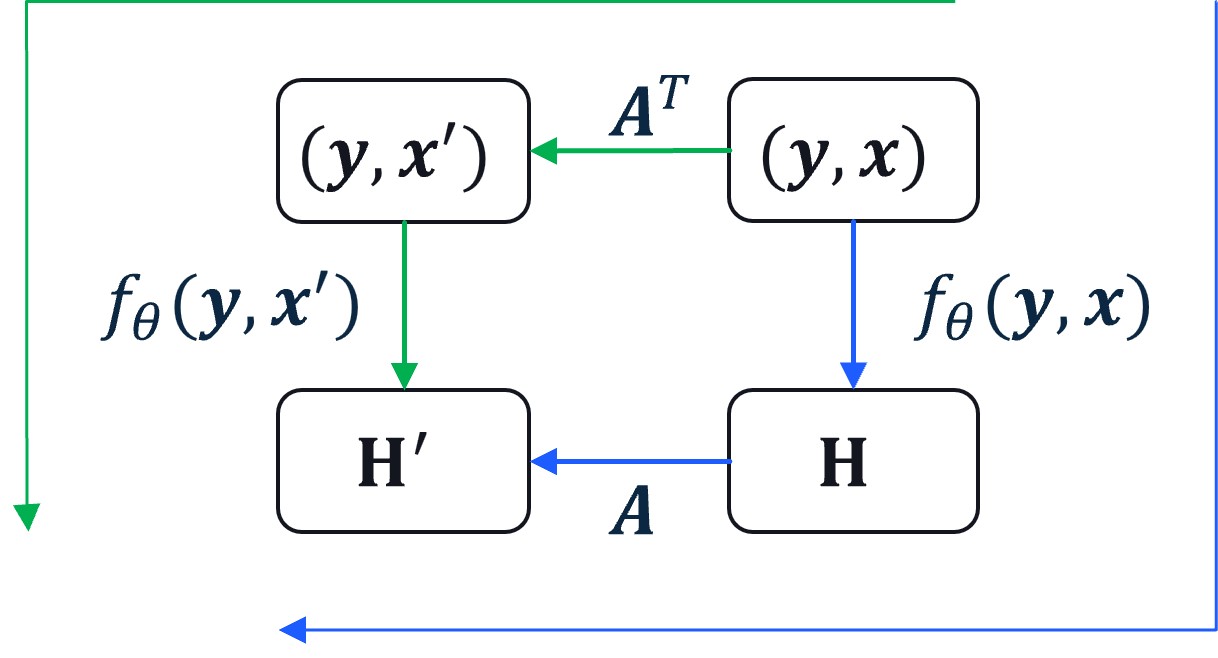}
    \caption{\textit{Flowchart illustrating the permutation equivariance property of ReQuestNet.} Whether permuting the DMRS vector before channel estimation or permuting the estimated channel matrix afterward, the result remains consistent. $\mathbf{H}$ represents the reconstructed channel matrix.}
    \label{fig:permutation_equivariance}
\end{figure}

Formally, let $\mathbf{\mathcal{A}}$ be a permutation matrix which when pre-multiplied or post-multiplied by a matrix, say $\mathbf{M}$, leads to permuting the rows or columns of $\mathbf{M}$ respectively. Define the permuted channel matrix as $\mathbf{H}' = \mathbf{H} \mathbf{\mathcal{A}}$ and the permuted pilot vector as $\mathbf{x}' = \mathbf{\mathcal{A}}^T \mathbf{x}$. The received signal remains unchanged:
\begin{equation}
    \mathbf{y} = \mathbf{H}'\mathbf{x}' + \mathbf{n} = \mathbf{H}\mathbf{\mathcal{A}}\mathbf{\mathcal{A}}^T \mathbf{x} + \mathbf{n} = \mathbf{Hx} + \mathbf{n}
    \label{eq:perm_equiv_full}
\end{equation}

This invariance implies that the CE model should satisfy:
\begin{equation}
    \mathbf{f}_{\theta}(\mathbf{y}, \mathbf{x}) \mathbf{\mathcal{A}} = \mathbf{f}_{\theta}(\mathbf{y}, \mathbf{\mathcal{A}}^T \mathbf{x}),
    \label{eq:perm_equiv}
\end{equation}
ensuring that permuting the input pilot vector results in a correspondingly permuted channel estimate. Rather than learning this symmetry from data, \sysname incorporates permutation equivariance as an architectural prior. This design choice improves learning efficiency, reduces overfitting, and enhances generalization across varying MIMO configurations. \sysname performs joint \ac{CE} across all spatial streams in a manner that is inherently equivariant to their ordering.

\begin{figure*}[!ht]
    \centering
    \includegraphics[width=\textwidth]{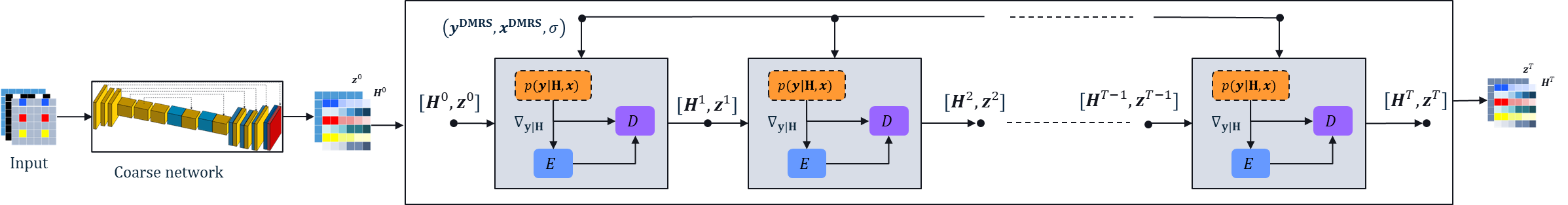}
    \caption{\textit{Overview of the ReQuestNet architecture.} ReQuestNet consists of a CoarseNet followed by a stack of refinement modules (RMs). Each gray-shaded unit represents an RM, which includes three sub-modules: the likelihood module (LM), the encoder module (E), and the decoder module (D). At each iteration, the encoder module updates the state variables based on information from the likelihood module \( p(\mathbf{y}|\mathbf{H}) \) and the preceding RM block. The decoder module then reconstructs the signal using the updated state variables, previous signal reconstruction, and information from the likelihood module.}
    \label{fig:requestnet_full}
    \vspace{-5mm}
\end{figure*}

\subsubsection{Modularity and Scalability}
While the previous discussion focused on estimating a single MIMO channel, practical wireless systems such as LTE, 5G NR, and Wi-Fi operate over a resource grid composed of multiple parallel MIMO channels. In 5G NR, for instance, the resource grid is structured into \acp{PRB}, each comprising a $12 \times 14$ frequency-time grid. These grids are subject to diverse configurations, including varying \ac{DMRS} patterns, \ac{PRB} grouping and precoding strategies (e.g., narrowband vs. wideband), dynamic transmit layer allocations, and multiple \ac{SCS} values.

A robust channel estimator must generalize across this wide range of configurations and channel conditions, including different delay profiles, mobility scenarios, \ac{SNR} levels, and spatial correlation structures. While classical statistical methods can be manually adapted to these variations by estimating channel statistics and tuning the estimator accordingly, this often incurs significant overhead and complexity. In contrast, \sysname is designed to be modular and scalable: it processes the resource grid in terms of fundamental units (e.g., PRBs per transmit layer) and incorporates specialized neural components to model interactions within and across these units. This design enables efficient adaptation to dynamic resource allocations and supports generalization across all supported 5G NR configurations, including variable RB counts, transmit layers, \ac{DMRS} patterns, and channel statistics such as delay spread, Doppler, \ac{SNR}, and MIMO correlation.

\subsubsection{Joint Channel Estimation across Resource Blocks}
To maximize processing gain, \sysname performs joint channel estimation across the entire resource grid. Traditional estimators such as linear \ac{MMSE} are limited in this regard, particularly under narrowband precoding where precoders vary across PRGs and are unknown to the receiver. In such cases, linear estimators treat differently precoded PRGs as uncorrelated, restricting the effective processing window to a single PRG.

\sysname overcomes this limitation by incorporating neural modules that implicitly estimate and compensate for precoder mismatches across PRGs. This enables the model to align and jointly process differently precoded sub-bands, effectively unlocking inter-PRG correlations and expanding the estimation window. This mechanism can be interpreted as a learned solution to a MRA problem discussed in Section~\ref{subsec:ce_mra}, akin to those encountered in cryo-electron microscopy \cite{Singer2018-le,Singer2020-af}. Joint estimation also improves robustness in high-delay spread environments. In 5G NR, DMRS symbols across antenna ports are generated using pseudo-random sequences derived from a length-31 Gold sequence \cite{Chen2021-ej}, allowing per-port estimation with a single symbol. However, in channels with high delay spread, linear de-patterning methods suffer from errors due to channel variation across adjacent DMRS tones. By jointly processing MIMO layers, \sysname mitigates these errors and improves estimation fidelity.

\section{ReQuestNet}\label{sec:requestnet-dl}

\subsection{ReQuestNet Architecture Overview}
We now present the architecture of ReQuestNet, designed in alignment with the guiding principles discussed earlier. Presented in Fig.~\ref{fig:requestnet_full}, \sysname implements a recurrent estimation framework that learns an iterative decoding algorithm, unrolled over $T+1$ steps. At each step, the network refines its channel estimate, progressively improving accuracy. Similar to RNNs, \sysname maintains auxiliary latent variables that serve as memory, encoding abstract representations of critical information such as inter-PRG precoder misalignment, cross-MIMO interference, and time-frequency selectivity.

The decoding process is divided into two stages: an initial step called \textit{CoarseNet}, followed by $T$ recurrent \textit{Refinement Steps}. CoarseNet performs per-PRG, per $T_x$–$R_x$ \ac{SISO} processing to produce locally contextualized state variables and coarse channel estimates. These outputs initialize the subsequent refinement steps, which operate jointly across the full resource grid.

Each of the $T$ refinement steps is implemented using a \textit{Refinement Module} (RM), a neural unit composed of three sub-modules: a \textit{Likelihood Module} (LM), an \textit{Encoder}, and a \textit{Decoder}. The LM injects feedback from the forward MIMO model by computing the gradient of the log-likelihood, while the encoder-decoder pair updates the latent state and channel estimate. Let $Z^t = \{\mathbf{z}_i^t\}_{i \in [N_{RB}]}$ and $\hat{H}^t = \{\hat{\mathbf{H}}_i^t\}_{i \in [N_{RB}]}$ denote the state variables and channel estimates at iteration $t$, respectively. Each RM takes $[Z^t, \hat{H}^t]$ as input and outputs updated values $[Z^{t+1}, \hat{H}^{t+1}]$ via:
\begin{equation}
    Z^{t+1}, \hat{H}^{t+1} = h_{\theta^{t+1}_{z, \text{H}}}\left(Z^{t}, \hat{H}^{t}, \nabla_{\mathbf{y}|\mathbf{H}} (\hat{H}^{t}) \right),
    \label{eq:joint_update}
\end{equation}
where $h_{\theta^{t+1}_{z, \text{H}}}$ is the RM operator and $\nabla_{\mathbf{y}|\mathbf{H}} (\hat{H}^{t})$ is the gradient of the log-likelihood term, providing feedback on the quality of the current estimate.

Each RM is designed to be permutation equivariant with respect to the ordering of MIMO streams and adaptable to dynamic wireless configurations. The encoder module updates the state variables $Z^t$ by modeling intra-PRG and inter-PRG dependencies, as well as cross-MIMO interactions, using transformer-based self-attention mechanisms. This facilitates information exchange across RBs and spatial streams, enabling each state $\mathbf{z}_{i,(j,k)}^t$ to accumulate the context necessary for refining $\hat{\mathbf{H}}_{i,(j,k)}^t$. The decoder is a pointwise SISO network that produces per-PRG, per $T_x$–$R_x$ channel estimates. It takes as input the current state $\mathbf{z}_{i,(j,k)}^t$, the previous estimate $\hat{\mathbf{H}}_{i,(j,k)}^{t-1}$, and the likelihood gradient, and outputs the refined estimate.

Inspired by \ac{RIM} \cite{Putzky2017-sq}, \sysname performs recurrent inference over \ac{DMRS} observations, using feedback from the forward model to iteratively refine its predictions. The update in \eqref{eq:joint_update} can be decomposed into two sequential steps, following the RE-MIMO framework \cite{Pratik2021-remimo}, for each $i \in [N_{PRG}]$ and $(j,k) \in [N_{Rx}] \times [N_{Tx}]$:
\begin{align}
    Z_{i,(j,k)}^{t+1} &= Z_{i,(j,k)}^{t} + h_{\theta^{t+1}_{e}}^{z} \left(Z_{i,(j,k)}^{t}, \hat{H}_{i,(j,k)}^{t}, \nabla_{\mathbf{y}|\mathbf{H}} (\hat{H}_{i,(j,k)}^{t}) \right), \label{eq:latent_upate} \\
    \hat{H}^{t+1}_{i,(j,k)} &= h_{\theta^{t+1}_{d}}^{\text{H}} \left(Z^{t+1}_{i,(j,k)}, \hat{H}^{t}_{i,(j,k)}, \nabla_{\mathbf{y}|\mathbf{H}}^{(j,k)} (\hat{H}^{t}_{i,(j,k)}) \right). \label{eq:signal_update}
\end{align}

Here, the first step updates the latent state using the previous state, channel estimate, and likelihood feedback—analogous to a gradient descent step. The second step refines the channel estimate, acting as a projection step, similar to the structure of algorithms such as iterative soft thresholding \cite{Behboodi2022-ls}. The initial values $Z_{i,(j,k)}^0$ and $\hat{H}_{i,(j,k)}^0$ are provided by CoarseNet and are subsequently refined through the stack of RMs.

\subsection{CoarseNet}\label{sec:CoarseNet}
\begin{figure}[!ht]
    \centering
    \includegraphics[width=0.98\linewidth]{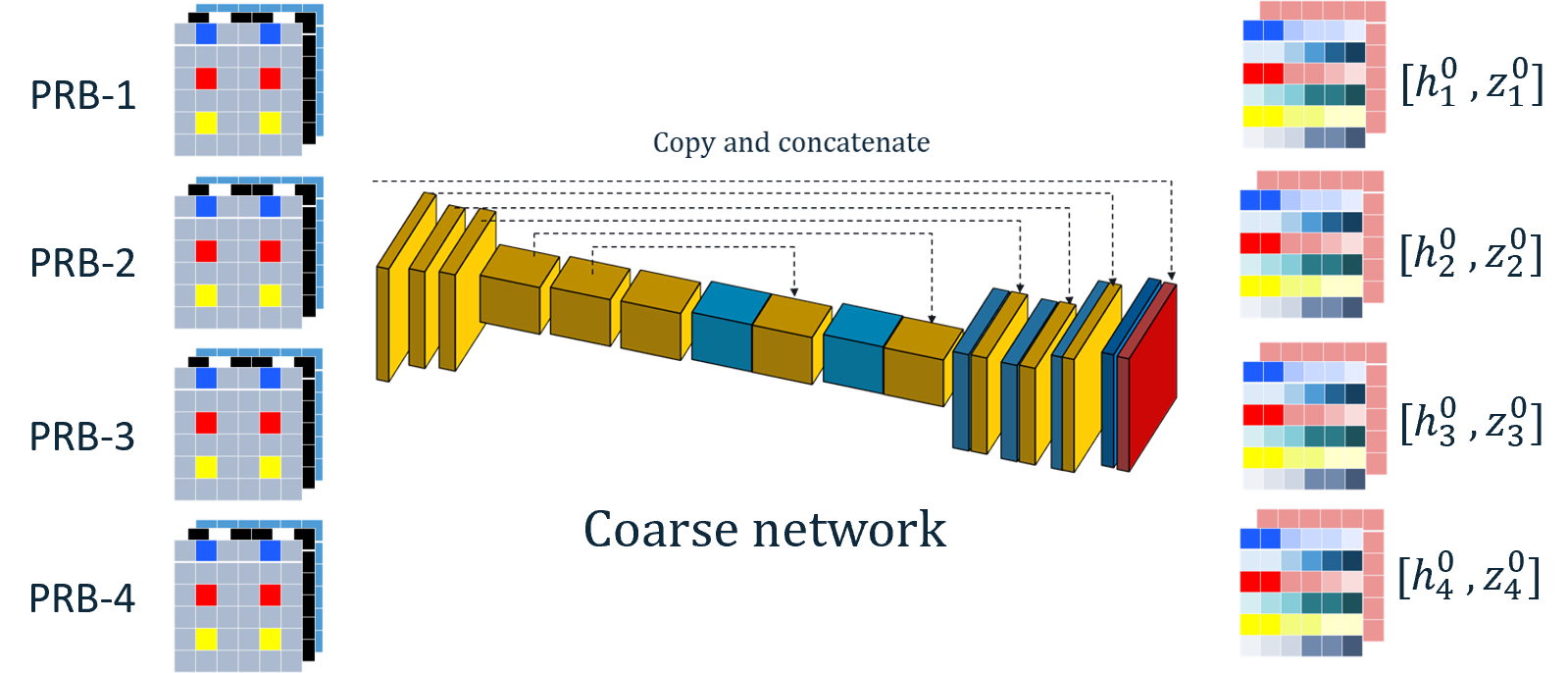}
    \caption{\textit{\coarsenn schematic.} \coarsenn performs per-PRG, per $T_x-R_x$ channel estimation based solely on the signal within the PRG sub-band.}
    \label{fig:coarsenet_2}
\end{figure}

\begin{figure*}[!ht]
    \centering
    \includegraphics[width=\linewidth]{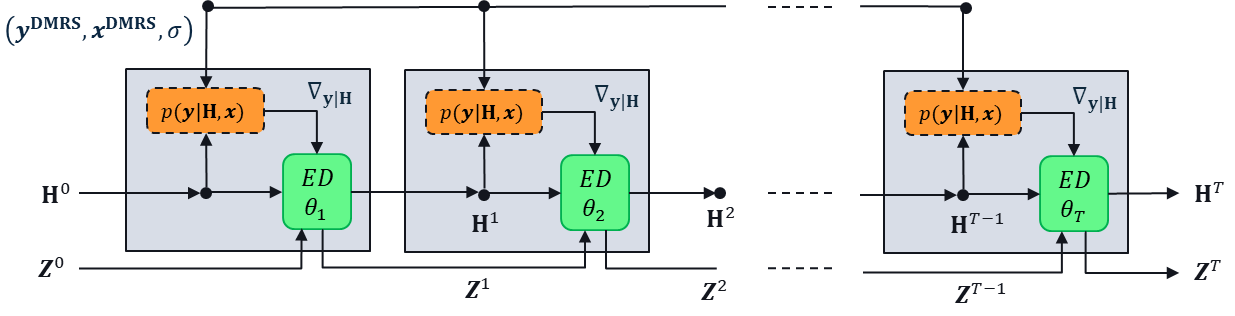}
    \caption{\textit{Temporal rollout of the RefinementNet framework.} The orange boxes represent the likelihood \( p(\mathbf{y|H}) \), while the green boxes denote the encoder-decoder (ED) modules. Each ED module is associated with a set of trainable parameters \( \theta_t \). The neural estimator iteratively reconstructs the transmit channel from the received signal using the gradient of the likelihood model \( p(\mathbf{y|H}) \) and the previous reconstruction.}
    \label{fig:refinementnet_1}
    \vspace{-5mm}
\end{figure*}

The first module in \sysname is CoarseNet, which performs initial channel estimation over the most localized sub-grid of the resource structure—specifically, per PRG and per $T_x$–$R_x$ pair as outlined in Fig.~\ref{fig:coarsenet_2}. This step serves as a non-linear, data-driven alternative to classical linear \ac{MMSE} estimation, relying solely on \ac{DMRS} observations. In contrast to MMSE, which requires explicit estimation of channel delay and Doppler profiles—often derived from auxiliary reference signals and prone to error propagation—\coarsenn learns to infer these characteristics implicitly from input data.

The input to \coarsenn is a multi-channel 2D grid corresponding to a single PRG. It includes: (i) the received signal at \ac{DMRS} locations, (ii) the transmitted \ac{DMRS} symbols, (iii) the LS channel estimate at \ac{DMRS} tones (computed as $\mathbf{y} \cdot \mathbf{x}^{\dagger}$), (iv) a binary mask $B$ indicating \ac{DMRS} positions, and (v) a noise standard deviation channel $\sigma_{snr}'$. Non-\ac{DMRS} REs are zero-padded in all relevant channels. The output consists of a bundle-wise SISO channel estimate and a set of locally contextualized latent variables used to initialize the refinement modules. The input-output relationship is defined as:
\begin{equation}
    Z^{0}_{i, (j,k)}, \hat{H}^{0}_{i, (j,k)} =  \mathrm{CoarseNet}_{\theta^{0}} \left(Y_{i,j}, X_{i,k}, Y_{i,j} \cdot X_{i,k}^{\dagger}, B, \sigma_{snr}'\right),
    \label{eq:coarse_nn_update}
\end{equation}
where $Y_{i,j}$ and $X_{i,k}$ denote the received and transmitted \ac{DMRS} symbols for the $j^{th}$ receive and $k^{th}$ transmit antennas in the $i^{th}$ PRG, for $i \in [N_{PRG}], j \in [N_{Rx}], k \in [N_{Tx}]$.

\textit{Module details:} \coarsenn is implemented as a 2D U-Net CNN, a widely adopted architecture in computer vision for tasks such as image inpainting and style transfer. To support multiple \ac{DMRS} patterns within a single model, standard convolutional layers are replaced with gated convolutional layers \cite{gated_cnn}, which improve adaptability to structured sparsity. The input tensor has shape $\left[N_r \cdot L \cdot M \cdot N, C_{in}^{c\text{-}net}, BS \cdot 12, 14\right]$, where convolution is applied over the last two dimensions. Here, $L$ denotes the number of allocated transmit layers, $M$ is the batch size, $N$ is the number of PRGs, and $BS$ is the PRG bundle size. The total number of RBs is given by $N_{RB} = N \times BS$. The number of input channels $C_{in}^{c\text{-}net}$ is $8$, corresponding to: $2$ channels each for the received \ac{DMRS} signal $Y_{i,j}$, the transmitted \ac{DMRS} symbols $X_{i,k}$, and the LS channel estimate $Y_{i,j} \cdot X_{i,k}^{\dagger}$ (each split into real and imaginary parts), and $1$ channel each for the binary DMRS mask $B$ and the noise standard deviation $\sigma_{snr}'$.

\subsection{RefinementNet}\label{sec:RefinementNet}
RefinementNet, illustrated in Fig.~\ref{fig:refinementnet_1}, performs recurrent refinement of the coarse channel estimates produced by CoarseNet. It facilitates information exchange across PRGs and spatial streams, enabling joint estimation over the full resource grid. \refinenn consists of a stack of $T$ RMs, each receiving the output of its predecessor as input. While all RMs share the same architecture, each has its own set of trainable parameters. The final output of \refinenn is taken from the last RM in the stack.

\begin{figure}[!ht]
    \centering
    \includegraphics[width=0.98\linewidth]{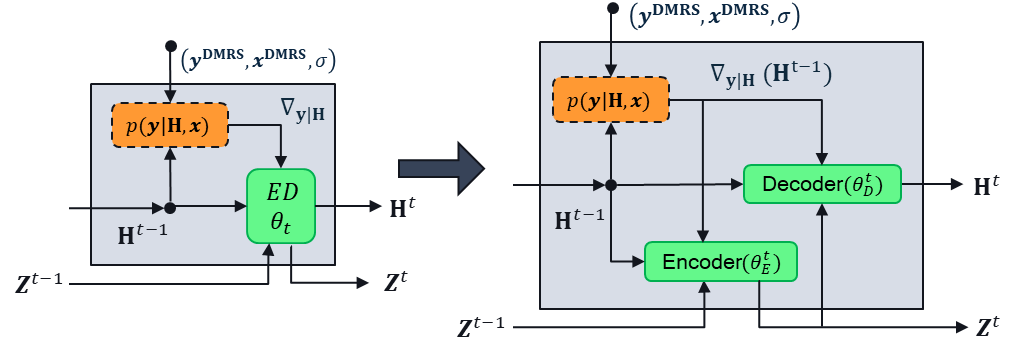}
    \caption{\textit{Detailed block description of one iteration step. The schematic on the right provides an in-depth illustration of the process depicted on the left.} $\theta^t_E$ and $\theta^t_D$ represent the weights associated with the $t^{th}$ encoder-decoder (ED) module. The encoder module jointly updates the state variables, which the decoder module then uses to produce predictions for each MIMO channel and PRG individually.}
    \label{fig:refinement_module}
    \vspace{-3mm}
\end{figure}

\begin{figure*}[!ht]
    \centering
    \includegraphics[width=\linewidth]{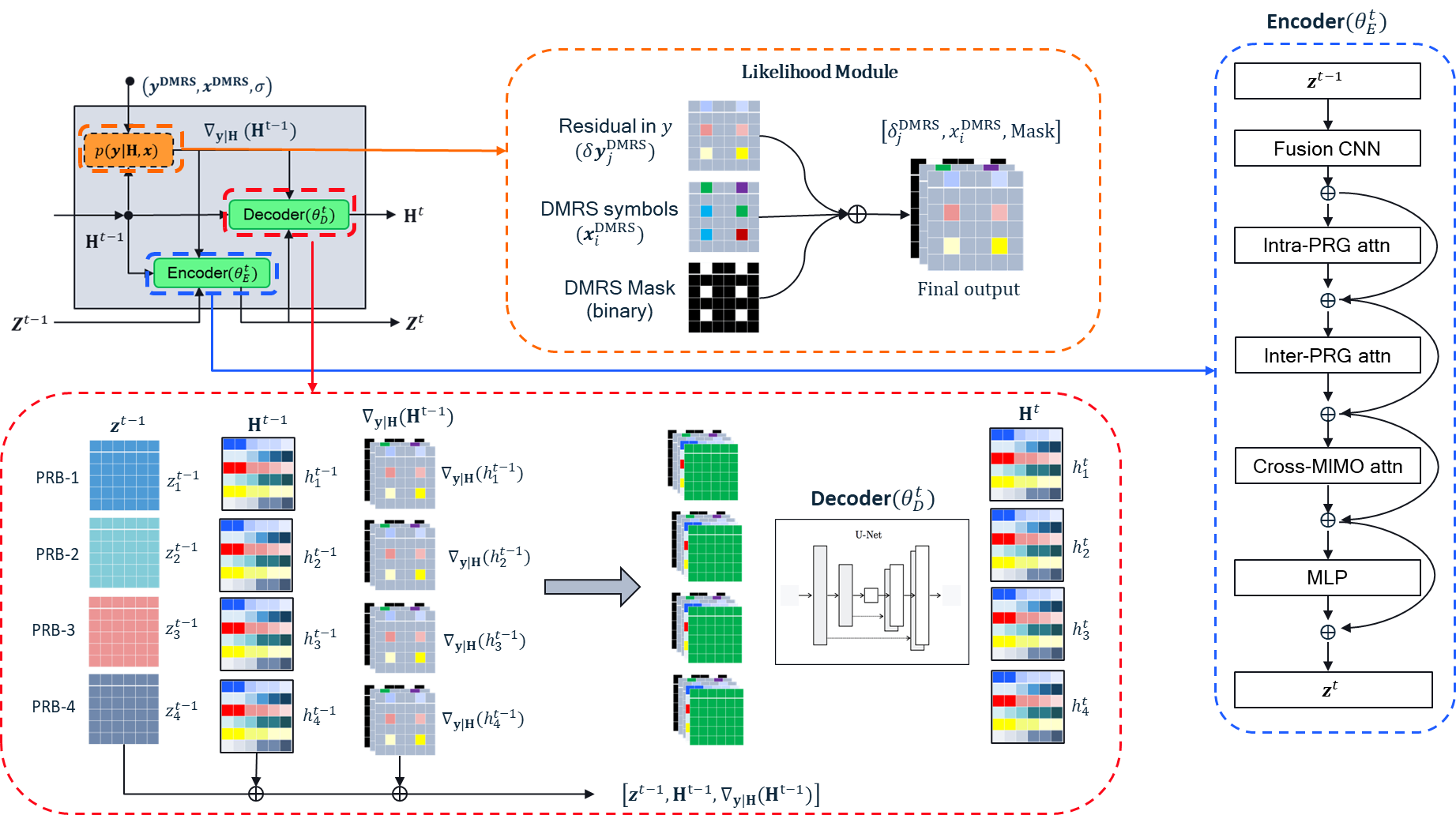}
    \caption{\textit{Schematic of the Refinement Module (RM)}. The likelihood module \( p(\mathbf{y|H}) \) (LM) computes gradient information as residual \( \delta \mathbf{y} \), DMRS mask, and DMRS symbols \( \mathbf{x} \), which are concatenated and forwarded to the encoder module. The encoder processes this input through a series of modules: Fusion CNN, Intra-PRG attention, Inter-PRG attention, Cross-MIMO attention, and MLP. The decoder, utilizing a U-Net architecture, reconstructs the channel by integrating the LM feedback, previous reconstruction, and encoder output.}
    \label{fig:refinementnet_schematic}
    \vspace{-5mm}
\end{figure*}

\subsection{Refinement Module}\label{sec:RefinementModule}

Each RM performs a single refinement step, analogous to one iteration of gradient descent. It is composed of three sub-modules: (i) the \textit{Likelihood Module} (LM), (ii) the \textit{Encoder Module}, and (iii) the \textit{Decoder Module}, as in Fig.~\ref{fig:refinement_module}. The LM injects feedback from the forward MIMO model, providing a signal that guides the refinement process. The encoder facilitates information exchange across PRGs and MIMO spatial dimensions, while the decoder updates the state variables and channel estimates based on the aggregated information.

\subsection{Likelihood Module}
The LM integrates knowledge of the MIMO generative model into the network by simulating the forward process and comparing it to the observed \ac{DMRS}. Specifically, it computes the gradient of the log-likelihood of the received signal under the current channel estimate:
\begin{equation}
    \nabla \log \text{p}(\mathbf{y}|\mathbf{H}, \mathbf{x}) \equiv \nabla_{\mathbf{y}|\mathbf{H}} = \frac{(\mathbf{y} - \mathbf{H}\mathbf{x}) \mathbf{x}^{\dagger}}{\sigma^2} = \frac{\delta \mathbf{y} \cdot \mathbf{x}^{\dagger}}{\sigma^2},
    \label{eq:lm_eq}
\end{equation}
where $\delta \mathbf{y} = \mathbf{y} - \mathbf{H}\mathbf{x}$ is the residual vector. This gradient serves as a feedback signal to the encoder and decoder modules.

The LM is designed to be permutation equivariant. Let $\mathcal{A}$ be a permutation matrix. Then, for permuted inputs $\mathbf{H}' = \mathbf{H}\mathcal{A}$ and $\mathbf{x}' = \mathcal{A}^T \mathbf{x}$, the gradient transforms as:
\begin{equation}
    \nabla_{\mathbf{y}|\mathbf{H}'}(\mathbf{y}, \mathbf{x}') = \frac{(\mathbf{y} - \mathbf{H} \mathbf{x}) (\mathbf{x}^{\dagger} \mathcal{A})}{\sigma^2} = \nabla_{\mathbf{y}|\mathbf{H}}(\mathbf{y}, \mathbf{x}) \mathcal{A},
    \label{eq:lm_eq_3}
\end{equation}
preserving the equivariance property.

Following RE-MIMO \cite{Pratik2021-remimo}, the LM computes per-stream gradients for each $j$–$k$ spatial stream:
\begin{equation}
    \nabla_{\mathbf{y}|\mathbf{H}}^{(j,k)} = \frac{\delta \mathbf{y}_j \cdot \mathbf{x}_k^{\dagger}}{\sigma^2},
    \label{eq:lm_eq_4}
\end{equation}
where $\delta \mathbf{y}_j$ and $\mathbf{x}_k$ are the $j^{th}$ and $k^{th}$ components of the residual and pilot vectors, respectively. The LM injects the gradient information in its decomposed form—$\delta \mathbf{y}_j$, $\mathbf{x}_k$, and $\sigma$—allowing the network to learn how best to utilize this information. Notably, $\delta \mathbf{y}_j$ and $\sigma$ are permutation invariant, while $\mathbf{x}_k$ is permutation equivariant. For computational efficiency, the encoder-decoder block omits $\sigma$ from its input, 
as it is already incorporated within the \coarsenn processing pipeline. 

\subsection{Encoder Module}
The encoder module, illustrated in Fig.~\ref{fig:refinementnet_schematic}, is responsible for modeling interactions across PRGs and MIMO spatial streams. It implements the update step in Eq.~\eqref{eq:latent_upate}, enriching the latent state variables with inter-PRG and cross-MIMO contextual information. To achieve this, the encoder is composed of specialized neural components, each designed to capture a distinct axis of correlation.

\paragraph{Fusion-CNN.} The Fusion-CNN integrates log-likelihood feedback into the latent representations. It is implemented as a pointwise U-Net 2D CNN operating per PRG and per $T_x$–$R_x$ stream. The input-output relation is:
\begin{equation*}
    Z^{t'}_{i, (j,k)} = \text{Fusion-CNN}_{\theta^{t}_{e}} \left(Z^{t}_{i,(j,k)}, \hat{H}^{t}_{i, (j,k)}, \nabla_{\mathbf{y}|\mathbf{H}}^{(j,k)} (\hat{H}^{t}_{i}) \right),
\end{equation*}
for all $i \in [N_{PRG}], (j,k) \in [N_{Rx}] \times [N_{Tx}]$. The output is flattened and passed through a sequence of three self-attention layers, each targeting a different axis of interaction.

\paragraph{Intra-PRG Attention.} This module captures dependencies among PRBs within a PRG. The input tensor is reshaped to $\left[N_r \cdot L \cdot M \cdot N, BS, E_z\right]$, where $E_z=168$ corresponds to the flattened per-RB embedding dimension ($12 \times 14$). Self-attention is applied along the PRB ($BS$) axis.

\paragraph{Inter-PRG Attention.} To model interactions across PRGs, the output of the intra-PRG attention is reshaped to $\left[BS, N_r \cdot L \cdot M, N, E_z\right]$. Mean pooling is applied along the $BS$ axis to obtain a single embedding per PRG, resulting in a tensor of shape $\left[N_r \cdot L \cdot M, N, E_z\right]$. Self-attention is then applied across the PRG axis ($N$). The resulting embeddings are broadcast back to match the original shape and added residually to the input.

\paragraph{Cross-MIMO Attention.} This module captures dependencies across MIMO spatial streams. The input is reshaped to $\left[M \cdot N \cdot BS, N_r \cdot L, E_z\right]$, and self-attention is applied along the MIMO axis ($N_r \cdot L$).

\paragraph{MLP.} A pointwise MLP is applied to each PRB and $T_x$–$R_x$ embedding. The input tensor is reshaped to $\left[N_r \cdot L \cdot M \cdot N \cdot BS, E_z\right]$, and the MLP consists of one hidden layer with dimensions $E_z \rightarrow 4E_z \rightarrow E_z$. Depicted in Fig.~\ref{fig:refinementnet_schematic}, each attention layer and the MLP is followed by a residual connection to facilitate gradient flow during training.

\subsection{Decoder Module}\label{sec:DecoderModule}

The decoder module, shown in Fig.~\ref{fig:refinementnet_schematic}, is a pointwise network that refines the channel estimate for each PRG and $T_x$–$R_x$ stream. Based on \eqref{eq:signal_update}, it takes as input the updated state variable, the log-likelihood gradient, and the previous channel estimate, and outputs the refined estimate.

\paragraph{Module details.} The decoder is implemented as a U-Net 2D CNN, similar to \coarsenn, and operates per PRG and per MIMO stream. The input tensor has shape $\left[N_r \cdot L \cdot M \cdot N, C_{in}^{dec}, BS \cdot 12, 14\right]$, where convolution is applied over the last two dimensions. The number of input channels $C_{in}^{dec}$ is $8$, comprising $1$ for $Z^{t+1}_{i,(j,k)}$, $2$ for $\hat{H}^{t}_{i, (j,k)}$ (real and imaginary parts), and $5$ for $\nabla_{\mathbf{y}|\mathbf{H}}^{(j,k)}$ (decomposed components). The decoder outputs the updated CE $\hat{H}^{t+1}_{i,(j,k)}$.

\section{Experiments} \label{sec:experiments}
\subsection{Model training details}
At the core of our training objective is the classical \ac{MSE} loss computed per decoding step \( t \), between the ground truth channel \( h^{\text{RE}}_{(j,k)} \) and the predicted channel \( \hat{h}^{\text{RE},t}_{(j,k)} \):
\begin{equation}
    \mathcal{L}_{t}^{\text{MSE}} = \frac{1}{|\text{REs}|} \sum_{\text{REs}} \sum_{{R_x}, T_x} \| h^{\text{RE}}_{(j,k)} - \hat{h}^{\text{RE},t}_{(j,k)} \|_{2}^{2},
    \label{eq:loss_mse}
\end{equation}
where \( |\text{REs}| \) denotes the total number of resource elements. Since the simulated channels are unit-powered \cite{mathworks_nrTDLChannel}, normalization by MIMO system size is unnecessary.

To encourage progressive refinement, we define the final loss as an SNR-scaled, weighted average of per-step losses:
\begin{equation}
    \mathcal{L} = \log_{10}\left[ \frac{1}{M} \sum_{m=1}^{M} \frac{1}{s_{m}} \sum_{t=0}^{T} w_{t} \cdot \mathcal{L}_{t, m}^{\text{MSE}} \right],
    \label{eq:loss_final}
\end{equation}
where \( M \) is the batch size, \( s_m \) is the SNR scaling factor for sample \( m \), and \( w_t \) is the weight assigned to decoding step \( t \), with \( \sum_t w_t = 1 \). We use linearly increasing weights \( w_t \propto t \) to emphasize later refinement steps, and scale losses by SNR in linear scale:
\begin{equation}
    w_t = \frac{2(t+1)}{(T+1)(T+2)}, \quad s_m = 10^{\text{SNR}/10} + 1.0
    \label{eq:loss_weights}
\end{equation}

To improve robustness against noise estimation errors and signal power fluctuations, we apply two data augmentations:

\begin{itemize}
    \item \textbf{SNR perturbation:} Additive uniform random noise \( \rho \sim \mathcal{U}(-5\text{dB}, +5\text{dB}) \) is applied to the genie SNR before forwarding it to the model.
    \item \textbf{Signal amplification:} The received signal \( \mathbf{y} \) is scaled by \( \sqrt{c} \), where \( c \sim \mathcal{U}(0.20, 5.0) \) is the randomly sampled power amplification factor, and the model is trained to predict the amplified channel \( \mathbf{H} \cdot \sqrt{c} \).
\end{itemize}

To compensate for the induced scaling in the loss, we adjust the SNR scaling factor as \( s_m \rightarrow s_m \cdot c \). Training is performed using online data generation, with simulation parameters sampled from the configuration space detailed in Table~\ref{table:sim_configs}. To simulate realistic transmission conditions during training, we adopt three precoding strategies across PRGs: \textit{SVD-based}, \textit{random}, and \textit{wideband}. In the \textbf{SVD-based} approach, we average the ground truth channel matrices over all REs within a PRG and perform singular value decomposition (SVD). The precoder is then formed as $\mathbf{U} \mathbf{V}^\dagger$, where $\mathbf{U}$ and $\mathbf{V}$ are the left and right singular vectors, respectively. The \textbf{random} strategy generates a $T_x \times L$ matrix per PRG, orthonormalized via Gram-Schmidt to ensure column-wise orthogonality. The \textbf{wideband} strategy applies identity precoding across PRGs, effectively bypassing spatial transformation. These configurations, listed under “Precoding Type” in Table~\ref{table:sim_configs}, builds ReQuestNet's robustness to diverse and dynamic precoding conditions.

\begin{figure*}[!ht]
    \centering
    \begin{subfigure}{0.32\textwidth}
        \centering
        \includegraphics[width=\linewidth]{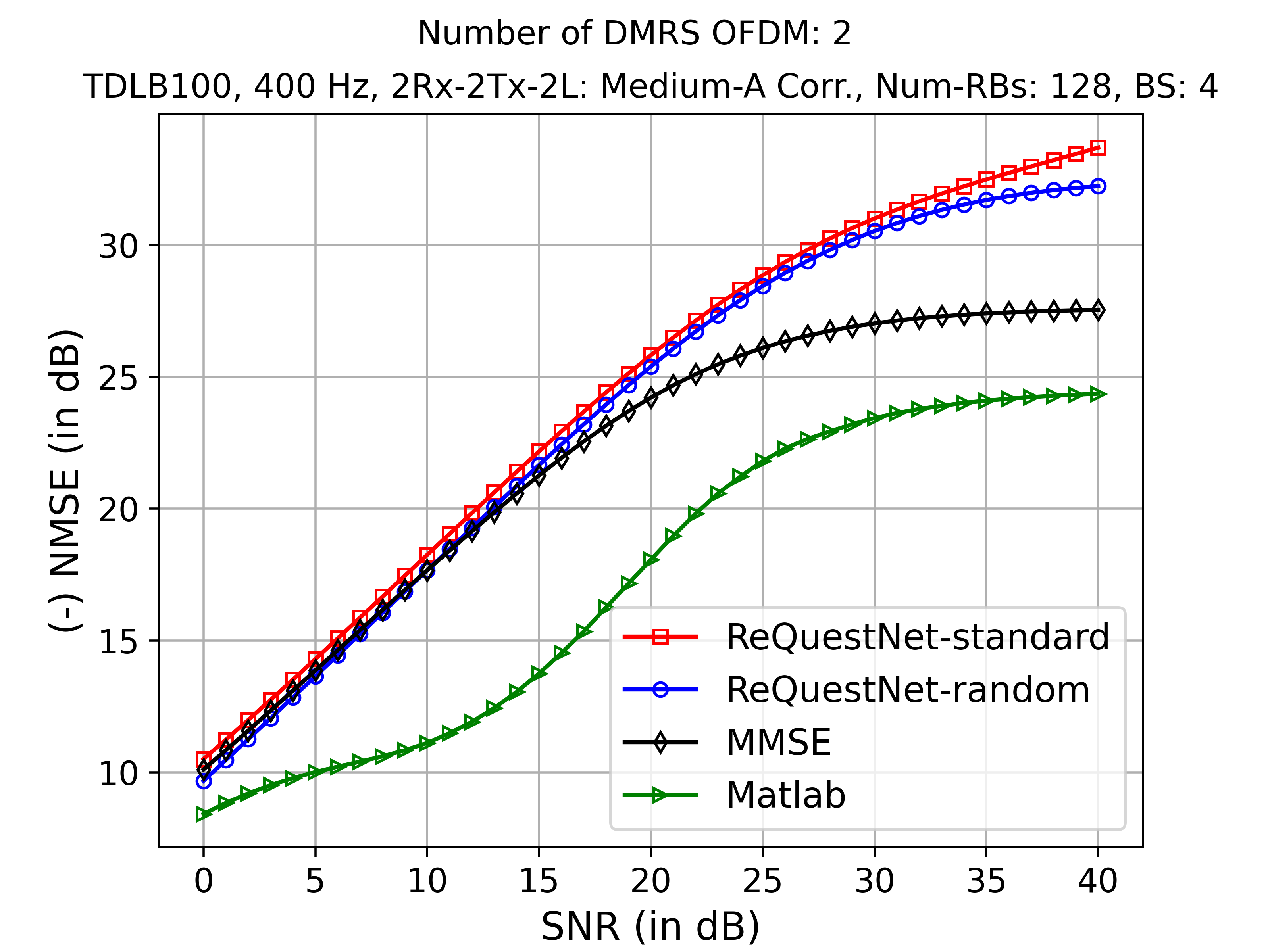}
        \label{fig:add_pos_1}
    \end{subfigure}
    \begin{subfigure}{0.32\textwidth}
        \centering
        \includegraphics[width=\linewidth]{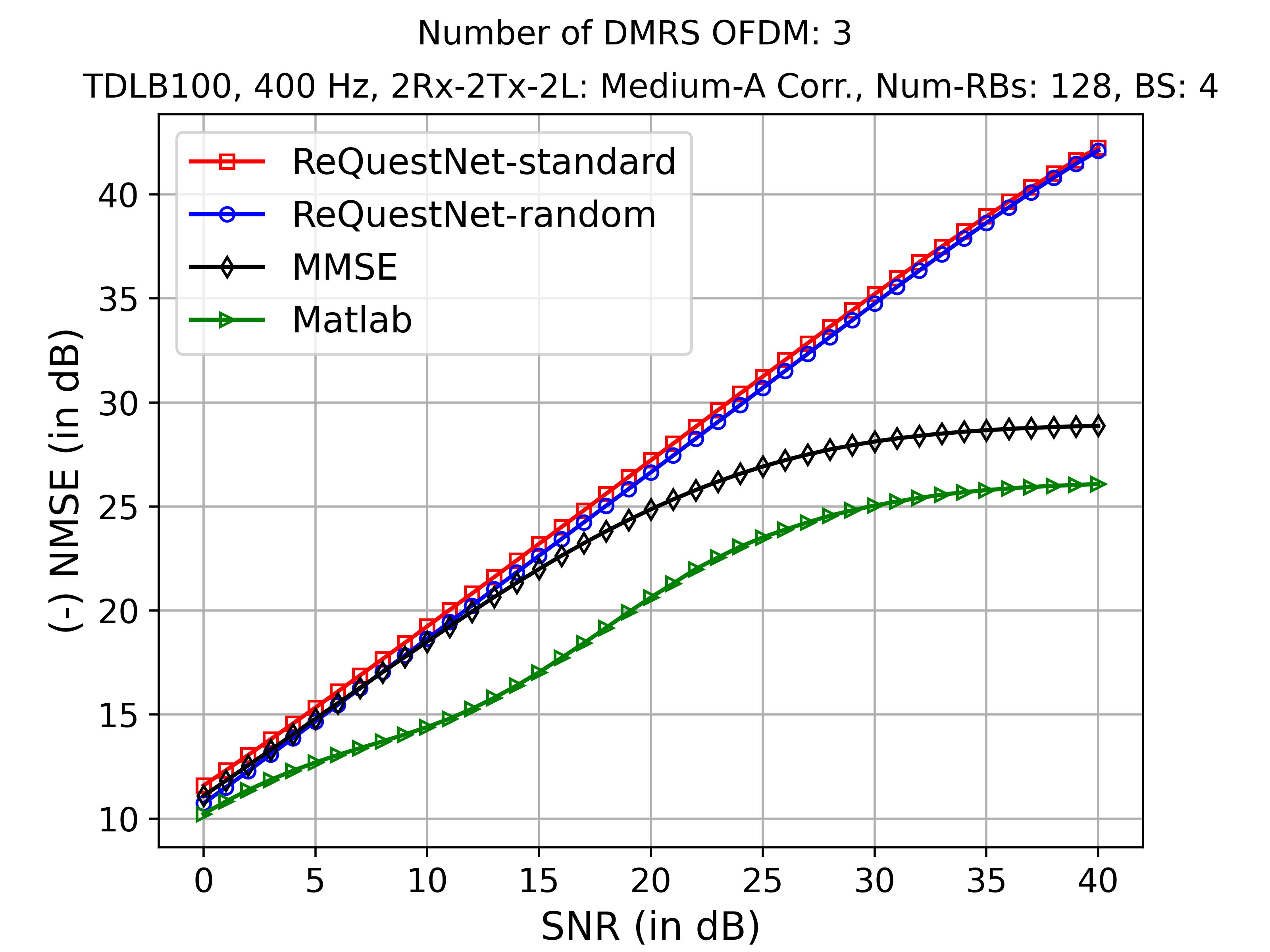}
        \label{fig:add_pos_2}
    \end{subfigure}
    \begin{subfigure}{0.32\textwidth}
        \centering
        \includegraphics[width=\linewidth]{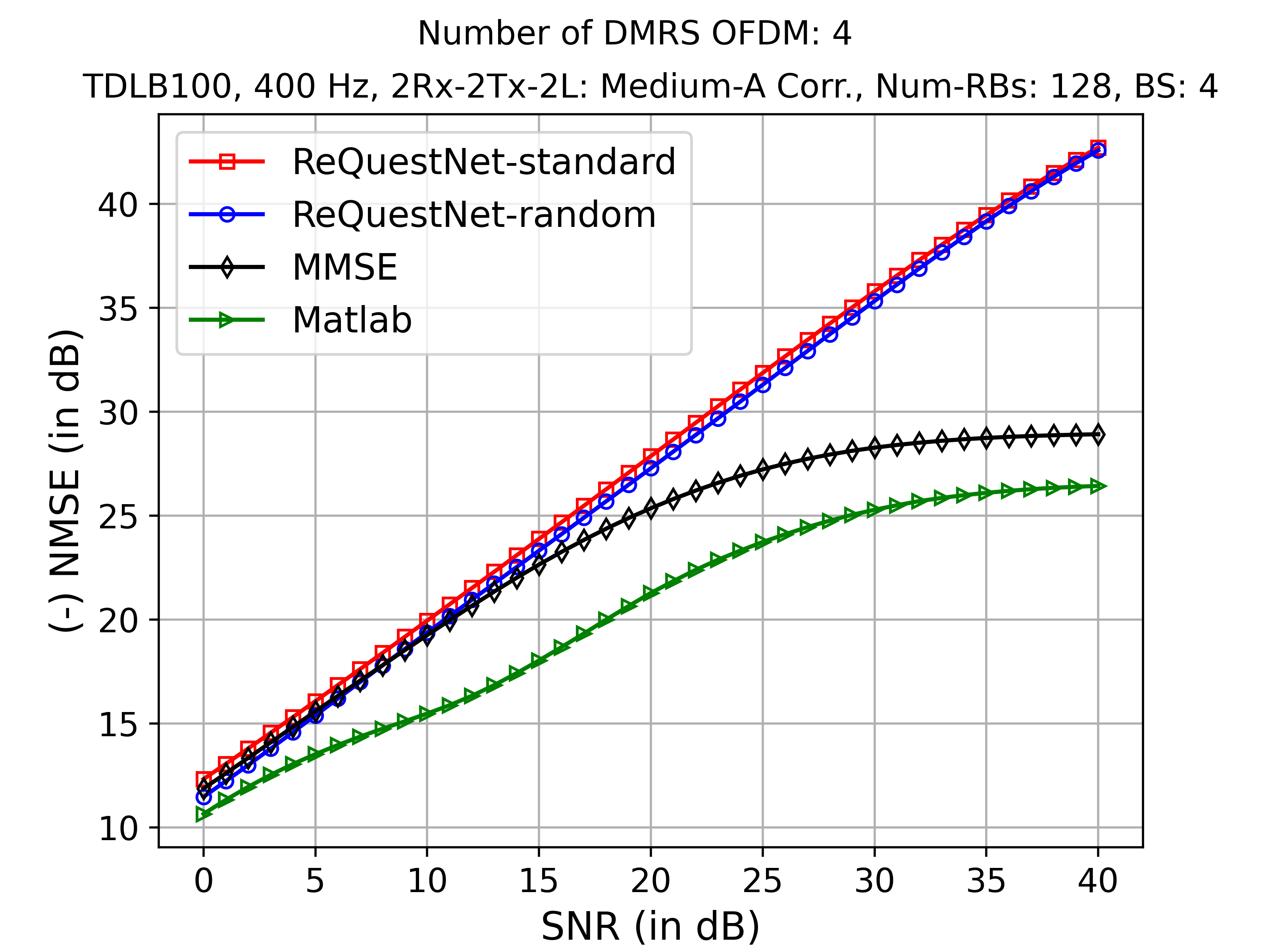}
        \label{fig:add_pos_3}
    \end{subfigure}
    \caption{($-$NMSE) vs. SNR for $2 \times 2$ MIMO CE for different values of number of DMRS OFDMs.}
    \label{fig:add_pos_plots}
    \vspace{-5mm}
\end{figure*}

\begin{table}[!ht]
\centering
\resizebox{\linewidth}{!}
{
\begin{tabular}{|>{\columncolor[gray]{0.8}}l||l|} 
\hline
\textbf{Simulation parameters} & \textbf{Possible values} \\ \hline
Delay profile (ReQuestNet \textit{standard})& TDLA, TDLB, TDLC, TDL-A30/B100/C300 \\ \hline
Delay spread & 1...300 ns \\ \hline
MIMO Correlation & Low, Medium, Medium-A, High \\ \hline
Sub-carrier spacing & 15, 30 kHz \\ \hline
Additional position & 1, 2, 3 \\ \hline
Configuration Type & 1, 2 \\ \hline
PRG Bundle size & 2, 4 \\ \hline
Max. Doppler shift & 0...450 Hz \\ \hline
Number of \ac{RB}s & 4...272 RBs \\ \hline
MIMO configuration ($T_x-R_x$) & $2\times2$ \\ \hline
Number of Layers & 1, 2 \\ \hline
SNR value & 0...40 dB \\ \hline
Precoding Type (per PRG bundle) & SVD, Random, Wideband \\ \hline
\end{tabular}
}
\caption{Comprehensive set of simulation parameters and their possible values used for training ReQuestNet.}
\label{table:sim_configs}
\end{table}

We train two variants of \sysname: \textit{random}, using custom \ac{TDL} profiles, and \textit{standard}, using standard \ac{TDL} profiles. For benchmarking, we compare against:

\begin{itemize}
    \item \textbf{Genie-aided MMSE:} An idealized MMSE estimator using perfect channel statistics (covariance matrices), perfect SNR, and no binning. Filter parameters are tailored per configuration and aligned with the PRG bundle size.
    \item \textbf{Matlab 5G Toolbox Estimator:} The \texttt{nrChannelEstimator} \cite{mathworks_nrChannelEstimate} performs pilot averaging and interpolation, supports TDL/CDL models, and incorporates noise estimation and delay profile approximation.
\end{itemize}

Performance is reported using the normalized mean squared error (NMSE):
\begin{equation}
    \text{NMSE} := \frac{1}{M} \sum_{m=1}^{M} \mathcal{L}_{t,m}^{\text{MSE}},
    \label{eq:nmse_def}
\end{equation}
where normalization by channel power is omitted due to the unit-power assumption. We use a batch size of 32 independent slots and train with the Adam~\cite{adam} optimizer for 40,000 steps, using a learning rate (lr) of \( 4 \times 10^{-4} \) and the \texttt{ReduceLROnPlateau} lr scheduler. The number of refinement steps in \refinenn is set to \( T = 4 \). \coarsenn{} comprises approximately $1.2\text{M}$ parameters, while each RM contains $0.6\text{M}$ parameters. Consequently, the total parameter count of \sysname{} amounts to $\sim\!3.6\text{M}$.

To comprehensively evaluate ReQuestNet, we categorize our experiments into three types:
\begin{itemize}
    \item \textbf{Modularity:} Tests the model's ability to handle variable input sizes and configurations, including different \ac{DMRS} patterns, PRG bundle sizes, and transmit layer allocations.
    \item \textbf{Standardization:} Benchmarks performance against classical CE methods across a range of standard channel and carrier configurations (e.g., delay spreads, Doppler shifts, MIMO correlations, delay profiles, and \ac{SCS} values).
    \item \textbf{Generalization:} Evaluates performance on unseen channels not seen during training.
\end{itemize}

All training and evaluation data are generated using the Matlab 5G Toolbox. For evaluation, we use the standard 3GPP channel models from \cite{3gpp_channel_models}.

\begin{table}[!ht]
\centering
\resizebox{\linewidth}{!}
{
\begin{tabular}{|>{\columncolor[gray]{0.8}}c||c|c|c|c|c| } 
\hline
Category & Num-DMRS & CT & BS & Num-Layers & Doppler \\
\hline
\textbf{Add-Pos} & \textbf{2}/\textbf{3}/\textbf{4} & 1 & 4  & 2 & 400 Hz\\ 
\hline
\textbf{Conf-Type} & 2 & \textbf{1}/\textbf{2} & 4 & 2 & 200 Hz\\
\hline
\textbf{Bundling} & 2 & 1 & \textbf{2}/\textbf{4} & 2 & 100 Hz\\
\hline
\textbf{Layers} & 2 & 1 & 4 & \textbf{1}/\textbf{2} & 100 Hz\\
\hline
\end{tabular}
}
\caption{Configurations for \textit{Modularity} experiments.}
\label{table:modularity_exp_parameters}
\end{table}

\subsection{Modularity Experiments}
This set of experiments evaluates ReQuestNet's ability to handle variably sized inputs and diverse 5G NR configurations, demonstrating its architectural flexibility and robustness.

\subsubsection{DMRS Additional Position}
As listed in Row 1 of Table.~\ref{table:modularity_exp_parameters} and visualized in Fig.~\ref{fig:add_pos_plots}, we evaluate performance across different numbers of DMRS symbols per slot. \sysname consistently outperforms MMSE across all configurations. Notably, the performance gap widens as the number of DMRS symbols goes from 2 to 3, highlighting ReQuestNet's ability to  leverage denser pilot patterns for improved estimation.

\subsubsection{DMRS Configuration Type}
\begin{figure}[!ht]
    \centering
    \begin{subfigure}{0.49\linewidth}
        \centering
        \includegraphics[width=\linewidth]{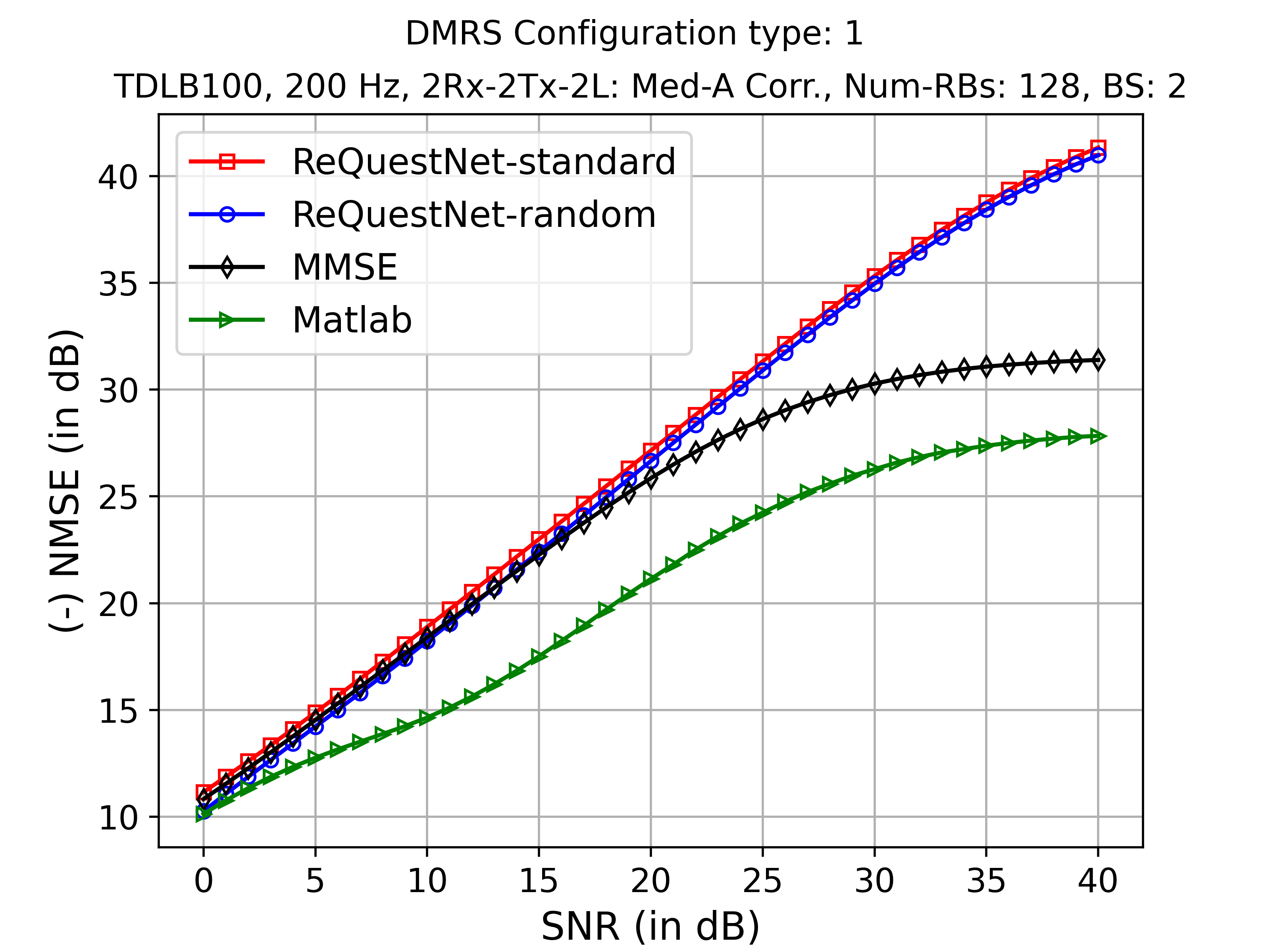}
        \label{fig:ct_1}
    \end{subfigure}
    \begin{subfigure}{0.49\linewidth}
        \centering
        \includegraphics[width=\linewidth]{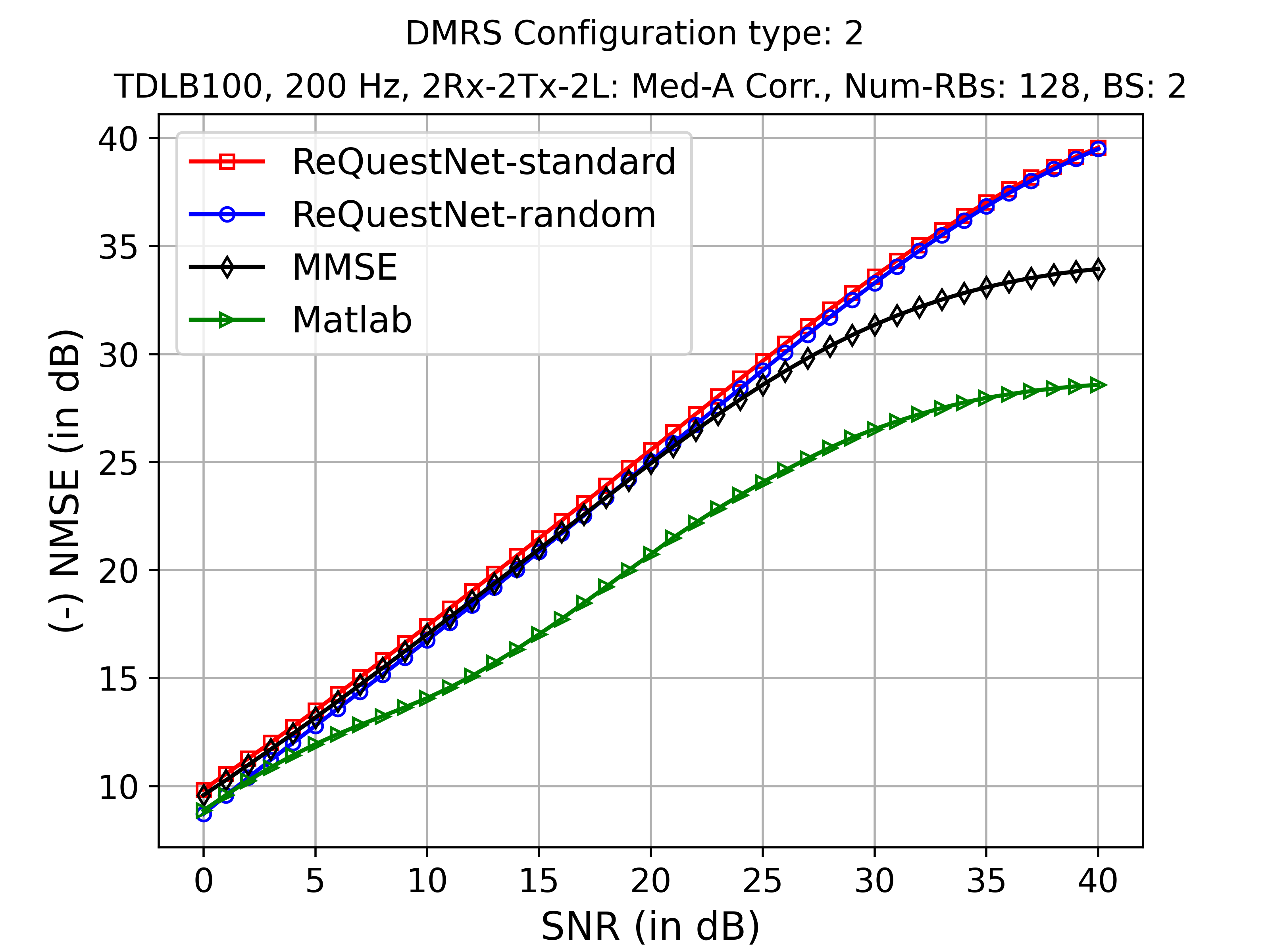}
        \label{fig:ct_2}
    \end{subfigure}
    \caption{($-$NMSE) vs. SNR for different DMRS configuration types.}
    \label{fig:config_type_plots}
\end{figure}

Shown in Row 2 of Table.~\ref{table:modularity_exp_parameters} and Fig.~\ref{fig:config_type_plots}, we compare performance across two DMRS configuration types (CTs). \sysname outperforms all baselines for both CTs, with a slightly larger margin observed for Configuration Type 1, indicating its adaptability to different pilot placement strategies.

\subsubsection{PRG Bundling}
\begin{figure}[!ht]
    \centering
    \begin{subfigure}{0.49\linewidth}
        \centering
        \includegraphics[width=\linewidth]{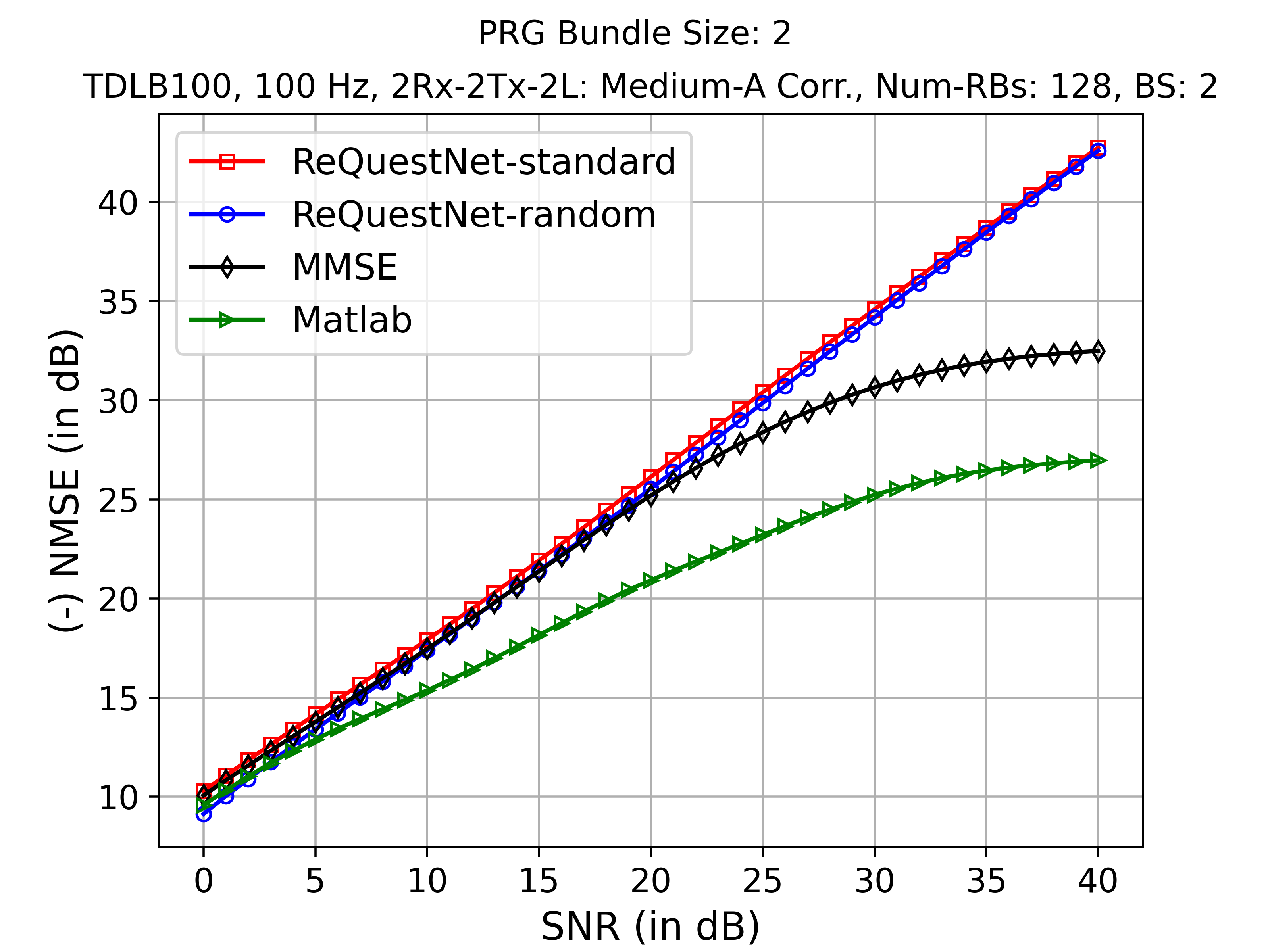}
        \label{fig:bs_2}
    \end{subfigure}
    \begin{subfigure}{0.49\linewidth}
        \centering
        \includegraphics[width=\linewidth]{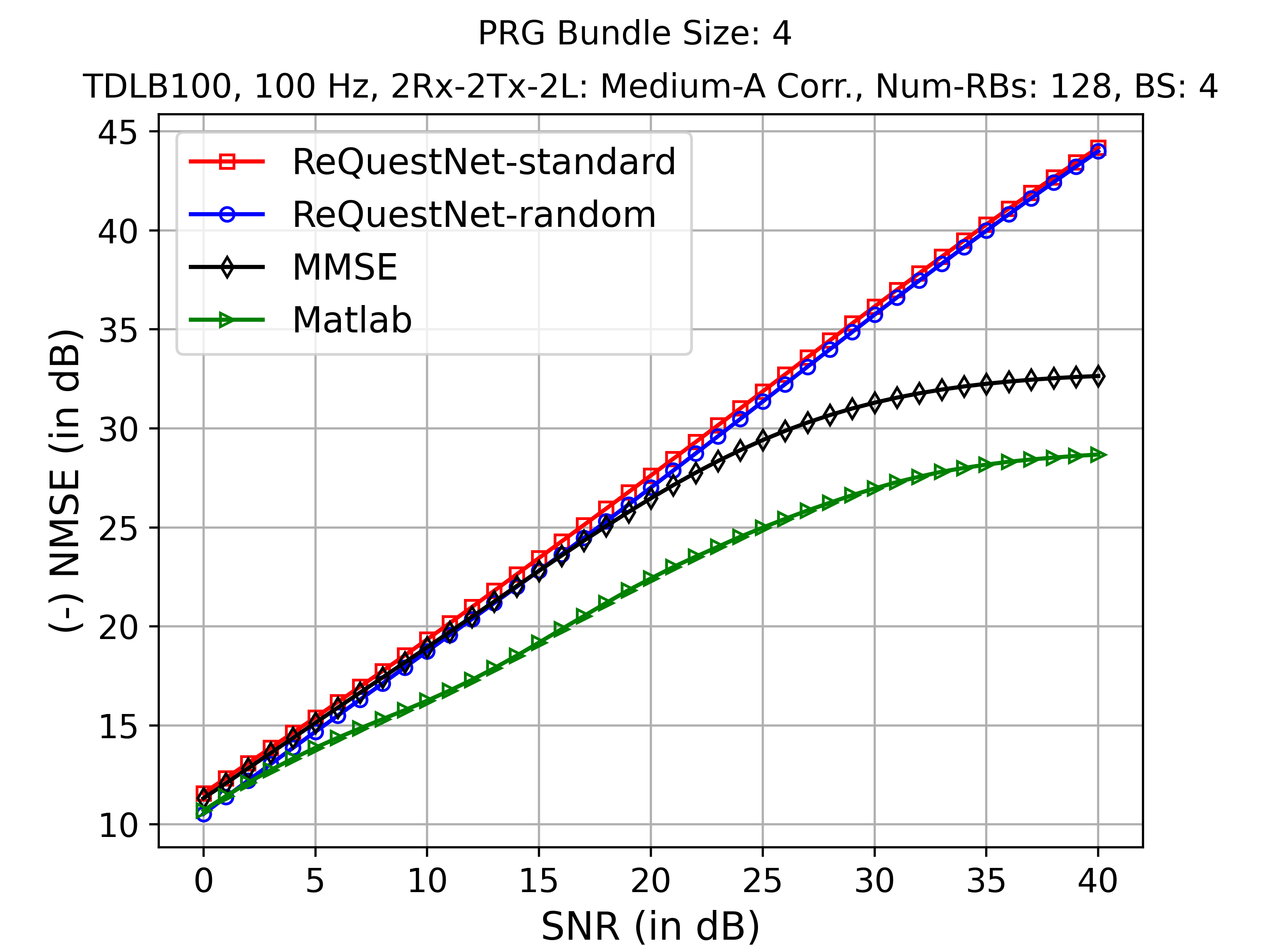}
        \label{fig:bs_4}
    \end{subfigure}
    \caption{($-$NMSE) vs. SNR for different PRG bundle sizes.}
    \label{fig:bundle_size_plots}
\end{figure}

In this experiment (Row 3, Table.~\ref{table:modularity_exp_parameters}), we assess performance under PRG bundle sizes of 2 and 4. As shown in Fig.~\ref{fig:bundle_size_plots}, \sysname maintains superior performance across both settings, demonstrating its ability to adapt to varying precoding granularities.

\subsubsection{Transmission Layers}
\begin{figure}[!ht]
    \centering
    \begin{subfigure}{0.49\linewidth}
        \centering
        \includegraphics[width=\linewidth]{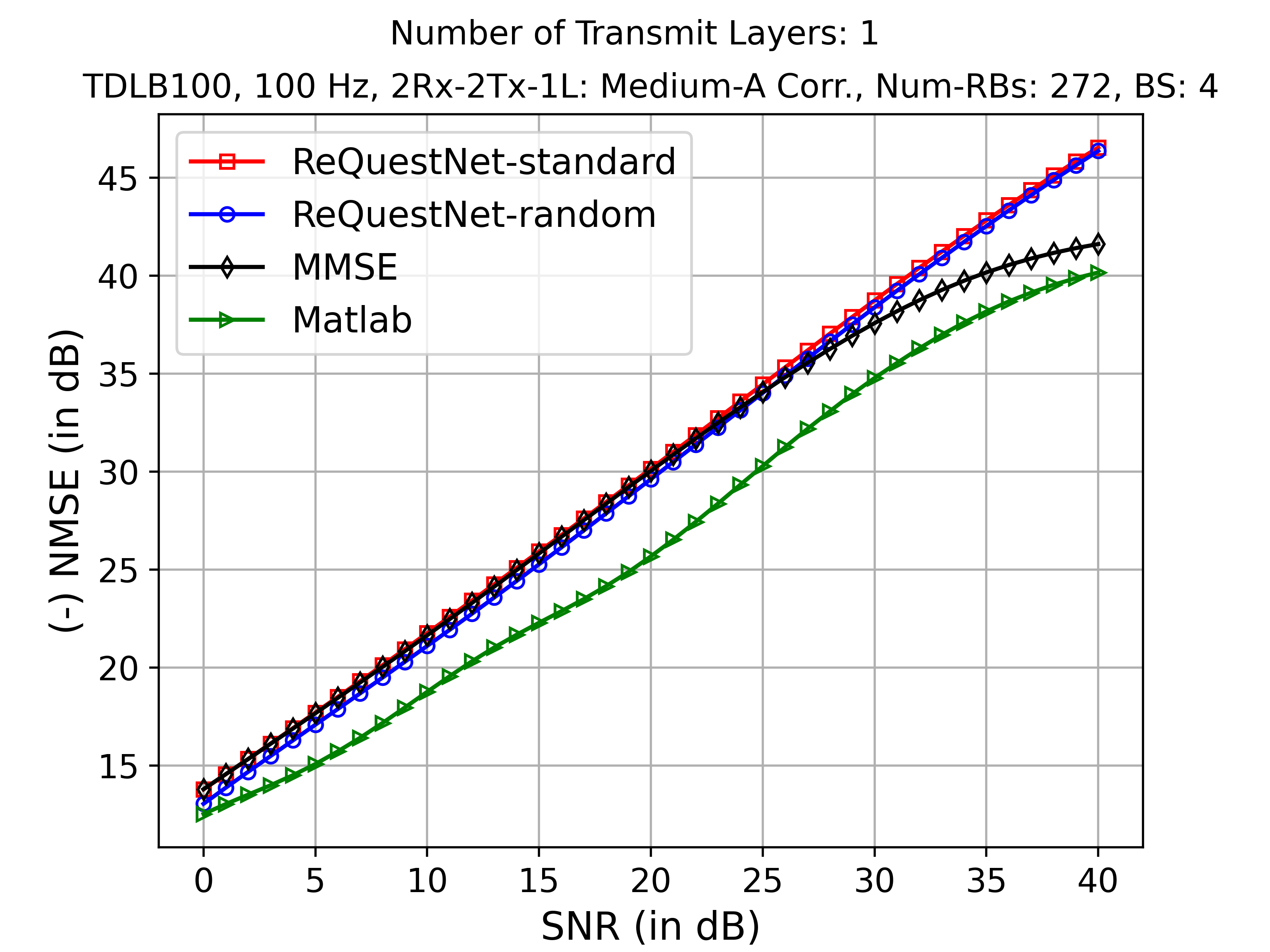}
        \label{fig:layer_1}
    \end{subfigure}
    \begin{subfigure}{0.49\linewidth}
        \centering
        \includegraphics[width=\linewidth]{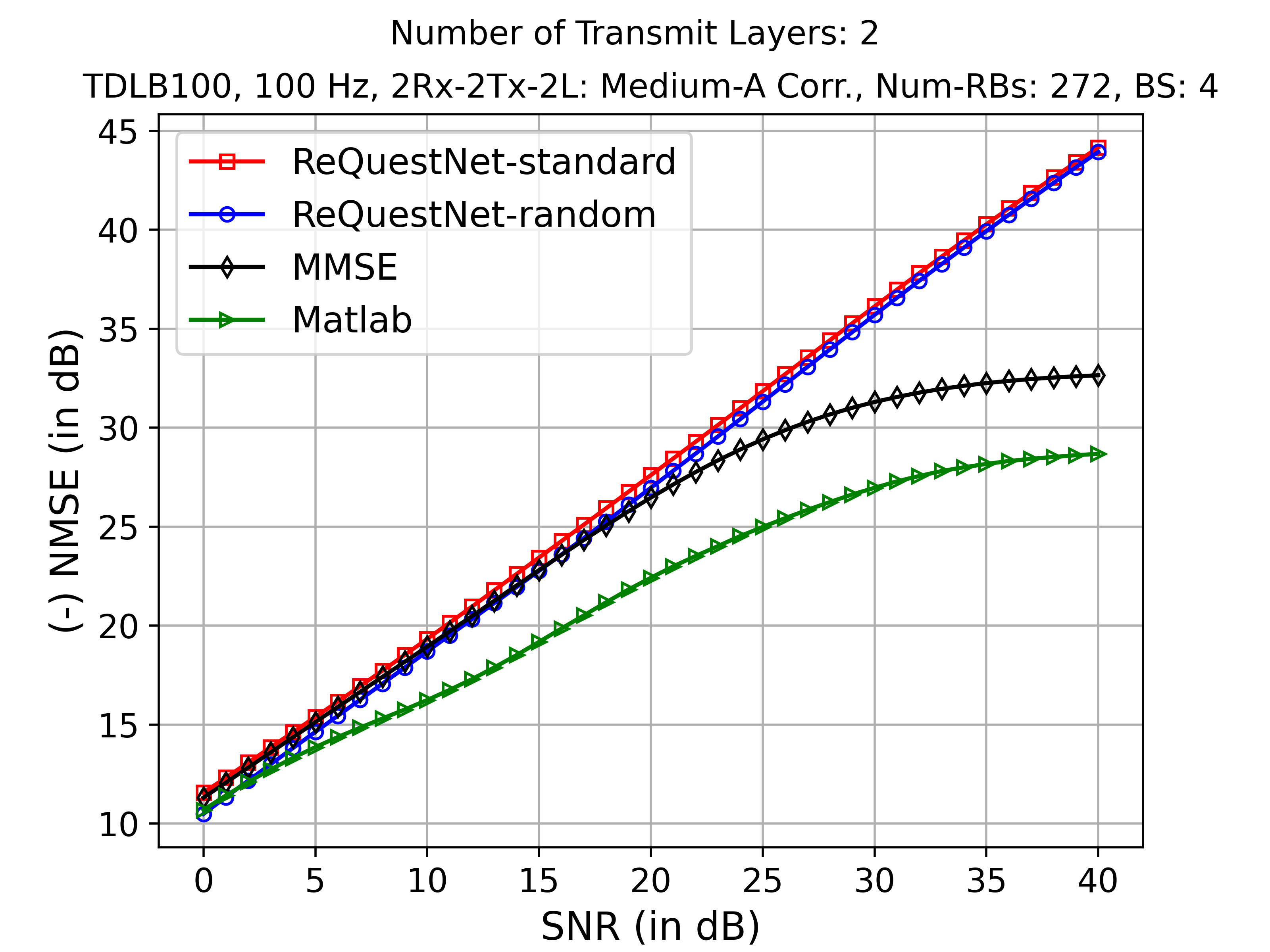}
        \label{fig:layer_2}
    \end{subfigure}
    \caption{($-$NMSE) vs. SNR for different for different number of $T_x$ layers for $2 \times 2$ MIMO.}
    \label{fig:num_layers_plots}
\end{figure}

As the number of transmit layers increases, CE becomes more challenging due to inter-layer interference. Leveraging its cross-MIMO attention mechanism, \sysname exhibits increasing performance gains over MMSE with higher layer counts, as shown in Fig.~\ref{fig:num_layers_plots} and detailed in Row 4 of Table.~\ref{table:modularity_exp_parameters}.

Unless otherwise specified, the base configuration for all modularity experiments includes: SCS = 30 kHz, MIMO = 2×2, Medium-A correlation, delay profile = TDL-B100, and Num-RBs = 128/272.

\begin{figure*}[!ht]
    \centering
    \begin{subfigure}{0.32\textwidth}
        \centering
        \includegraphics[width=\linewidth]{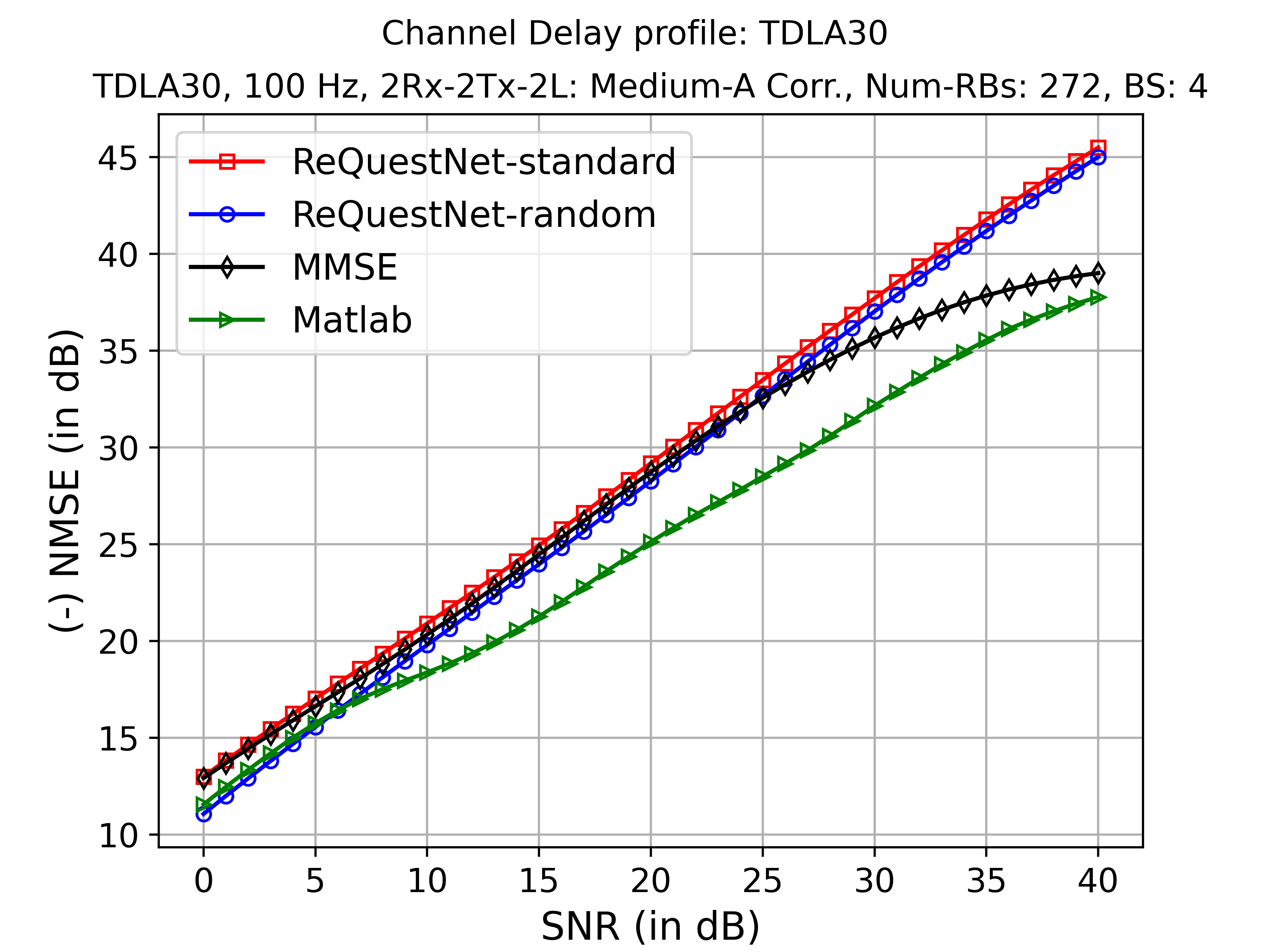}
        \label{fig:tdla30}
    \end{subfigure}
    \begin{subfigure}{0.32\textwidth}
        \centering
        \includegraphics[width=\linewidth]{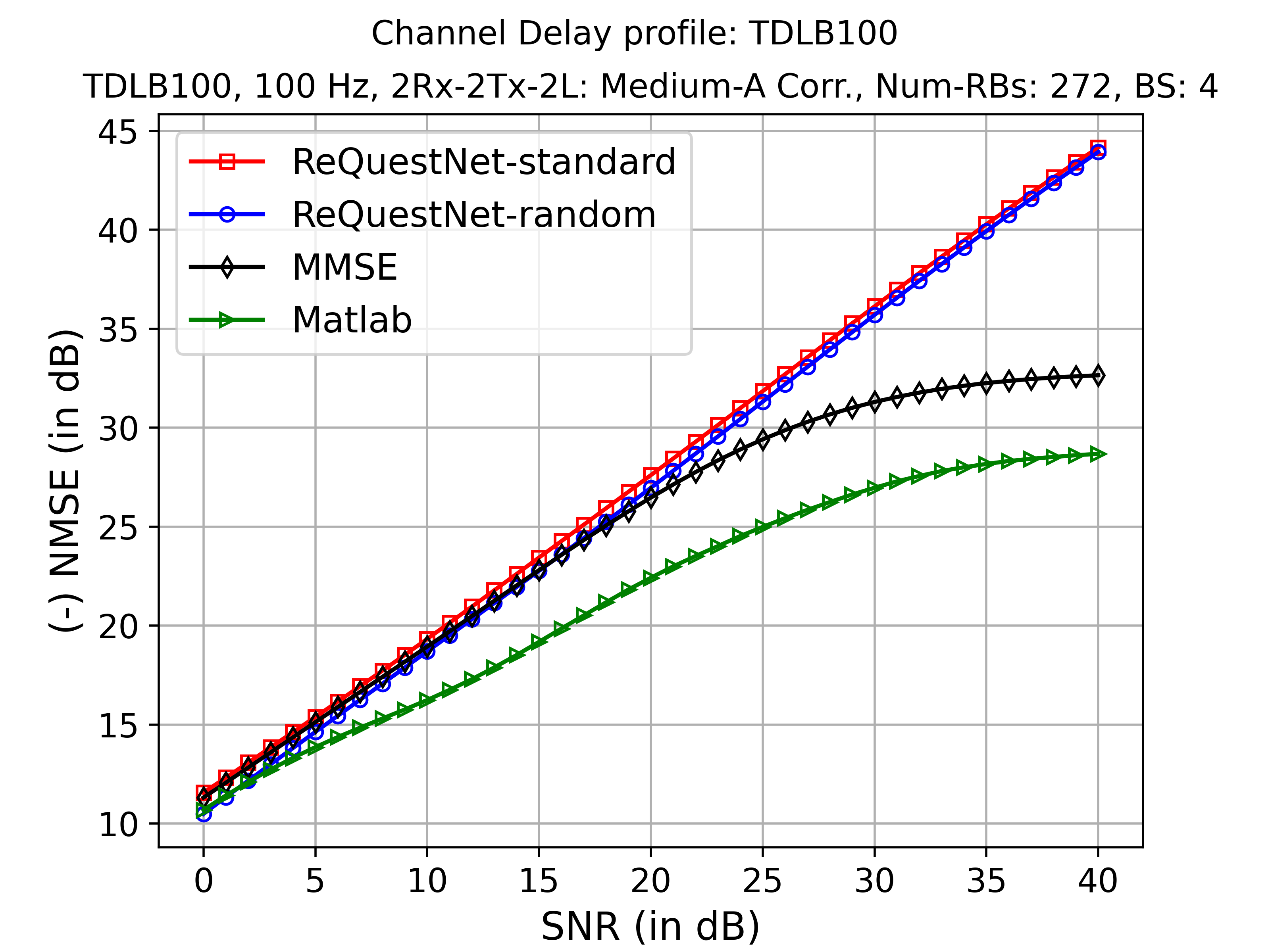}
        \label{fig:tdlb100}
    \end{subfigure}
    \begin{subfigure}{0.32\textwidth}
        \centering
        \includegraphics[width=\linewidth]{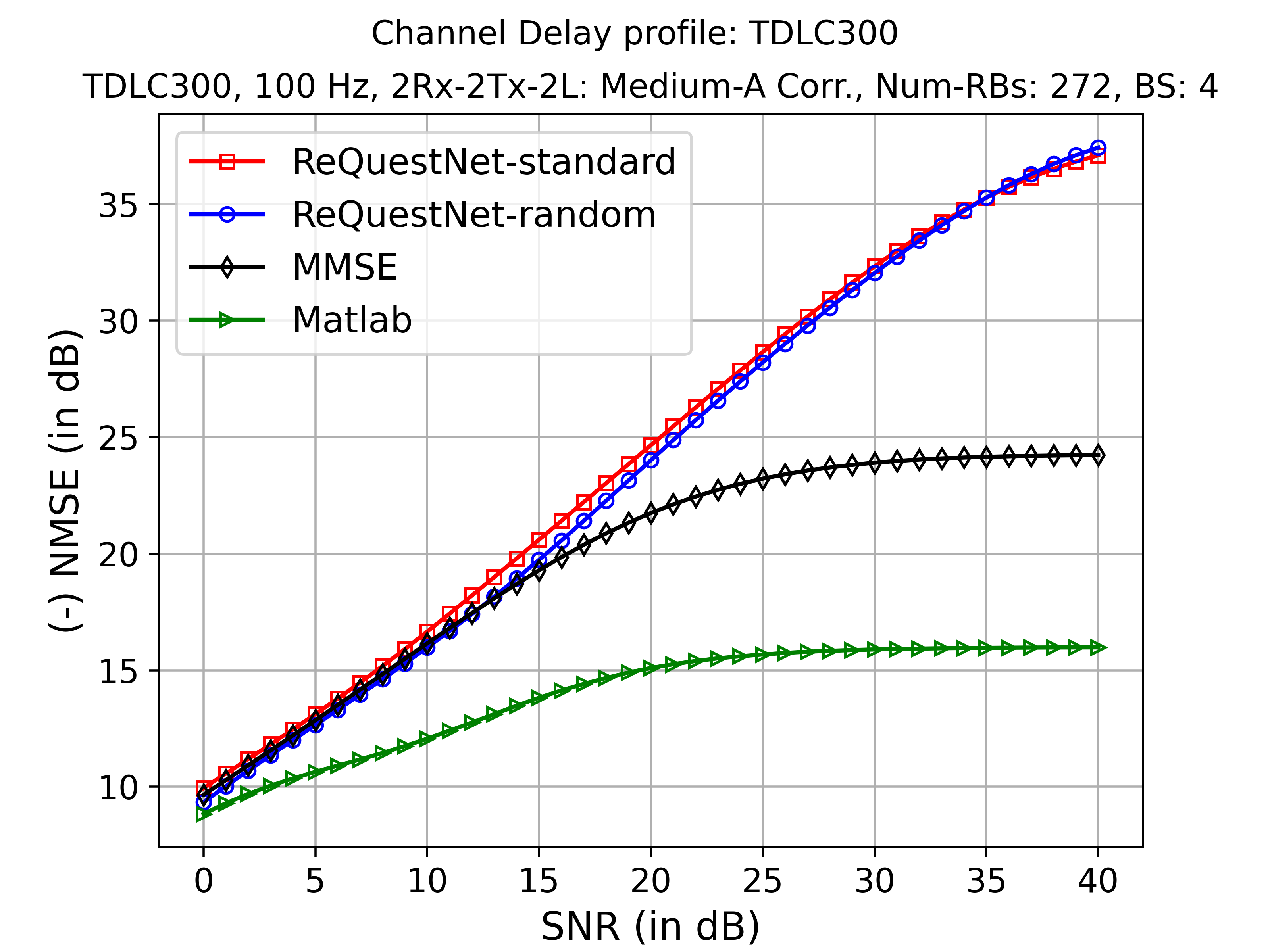}
        \label{fig:tdlc300}
    \end{subfigure}
    \caption{($-$NMSE) vs. SNR for different channel delay profiles and delay spread values.}
    \label{fig:delay_profiles_plots}
    \vspace{-3mm}
\end{figure*}

\begin{figure*}[!ht]
    \centering
    \begin{subfigure}{0.245\textwidth}
        \centering
        \includegraphics[width=\linewidth]{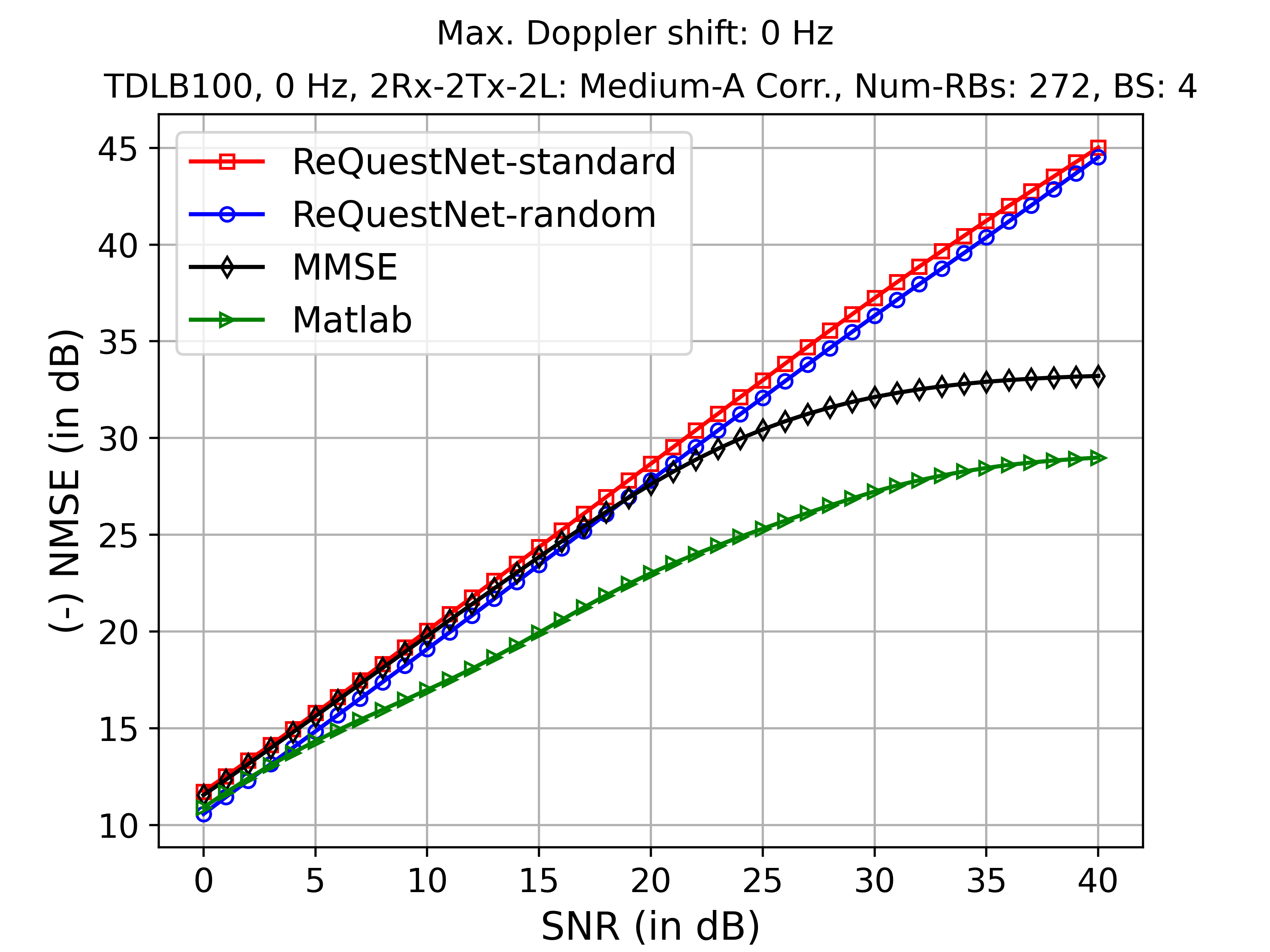}
        \label{fig:doppler_0}
    \end{subfigure}
    \begin{subfigure}{0.245\textwidth}
        \centering
        \includegraphics[width=\linewidth]{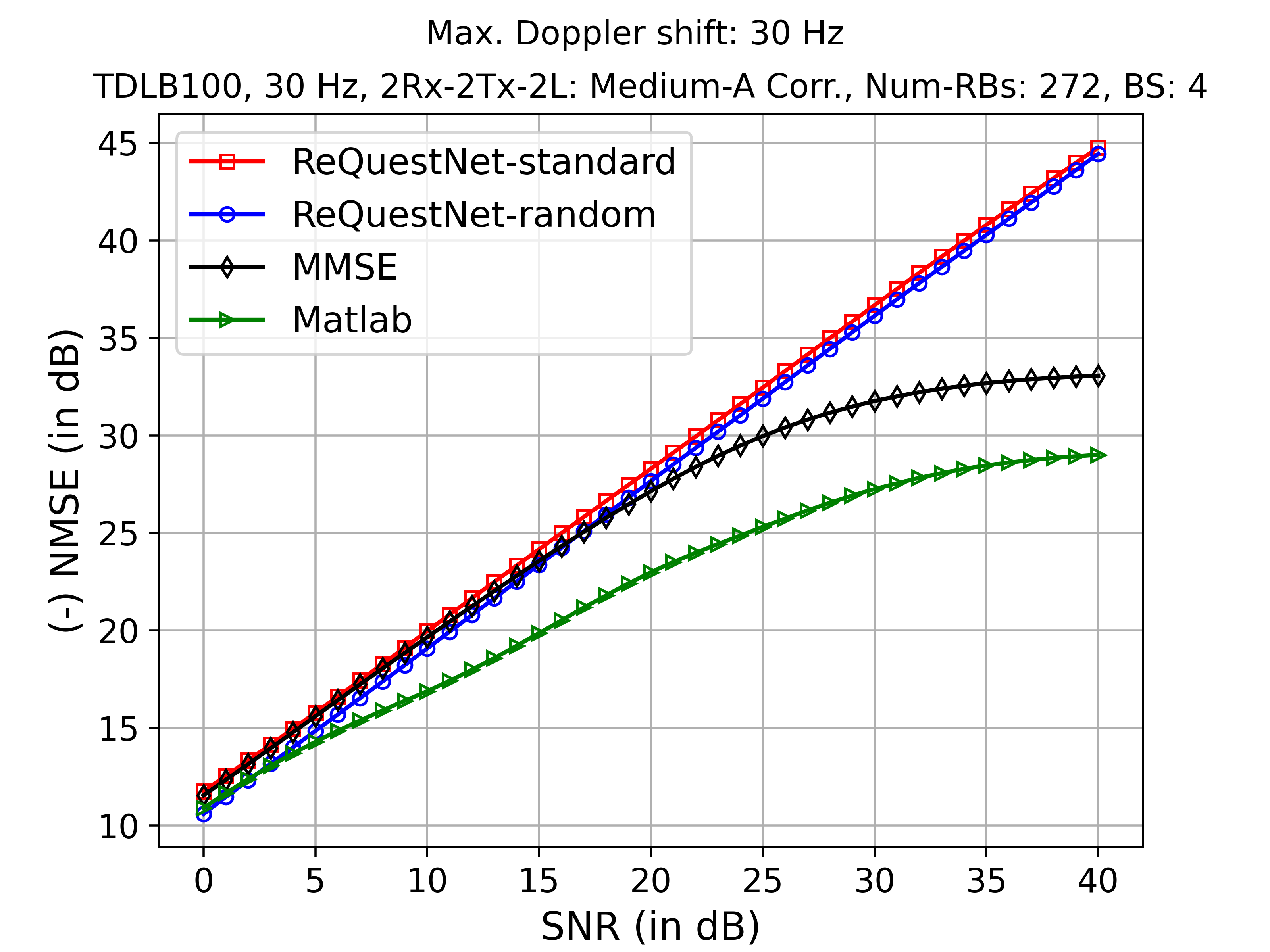}
        \label{fig:doppler_30}
    \end{subfigure}
    \begin{subfigure}{0.245\textwidth}
        \centering
        \includegraphics[width=\linewidth]{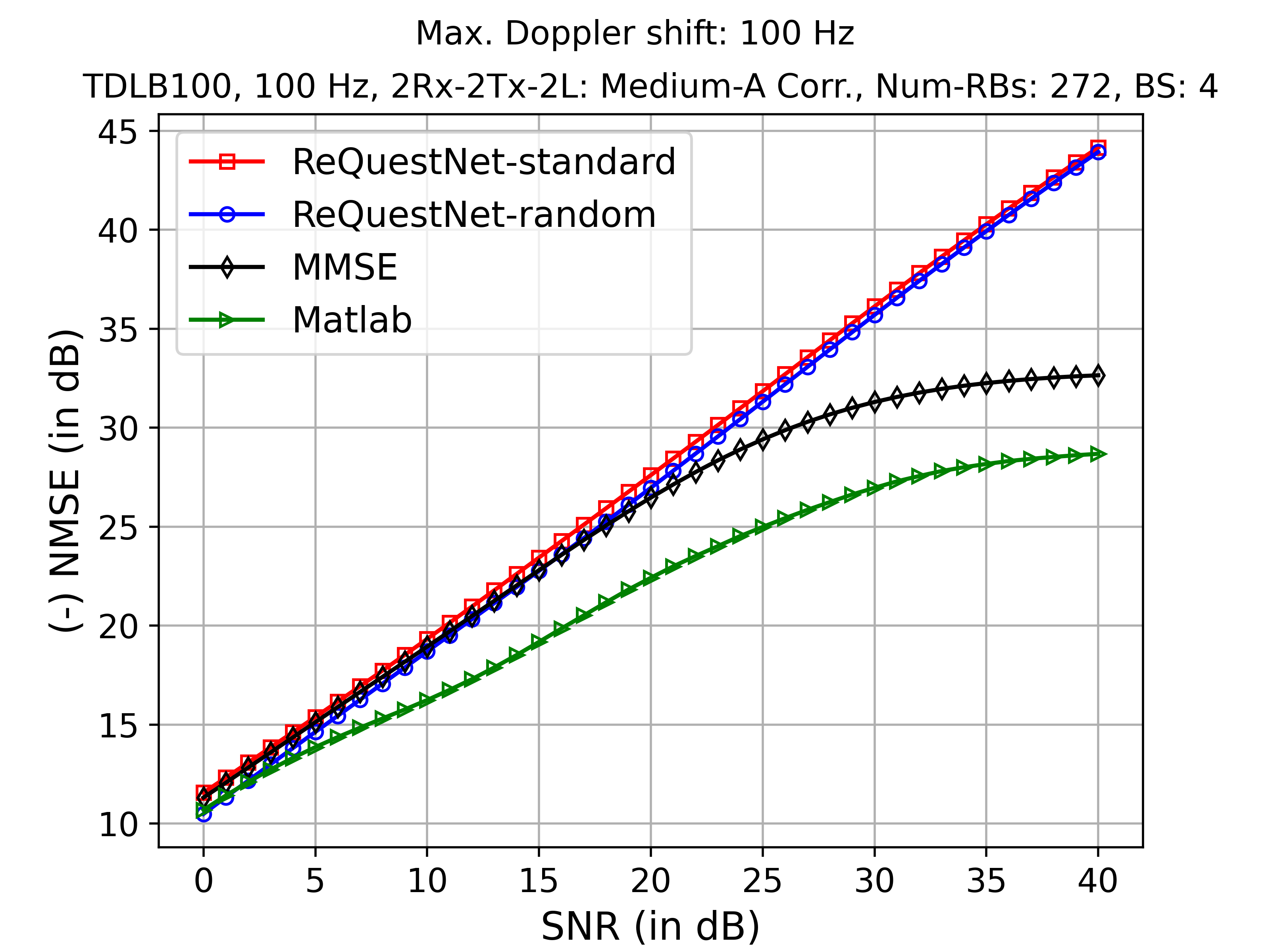}
        \label{fig:doppler_100}
    \end{subfigure}
        \begin{subfigure}{0.245\textwidth}
        \centering
        \includegraphics[width=\linewidth]{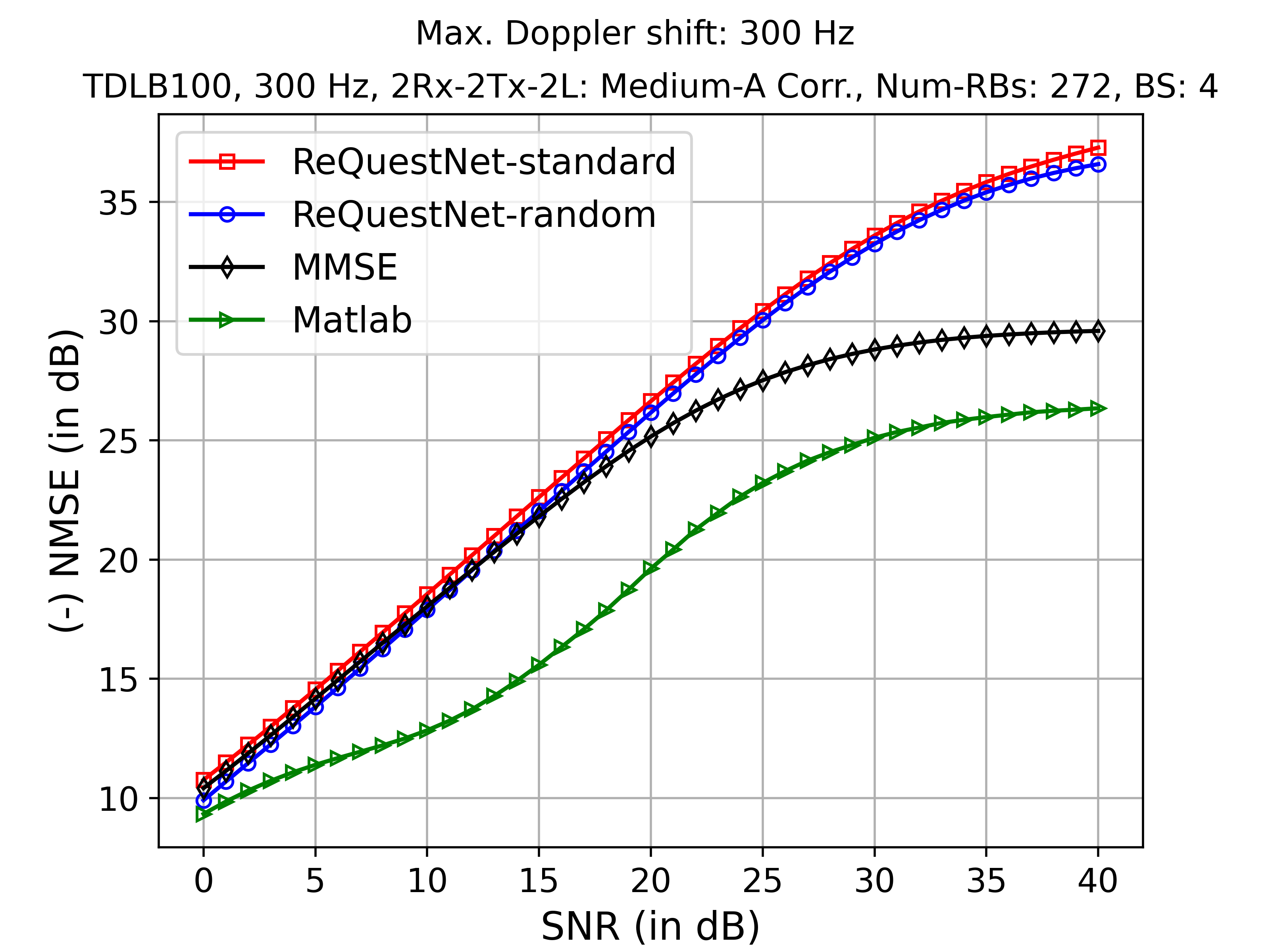}
        \label{fig:doppler_300}
    \end{subfigure}
    \caption{($-$NMSE) vs. SNR for different Max. Doppler shift values.}
    \label{fig:doppler_plots}
\end{figure*}

\begin{figure*}[!ht]
    \centering
    \begin{subfigure}{0.245\textwidth}
        \centering
        \includegraphics[width=\linewidth]{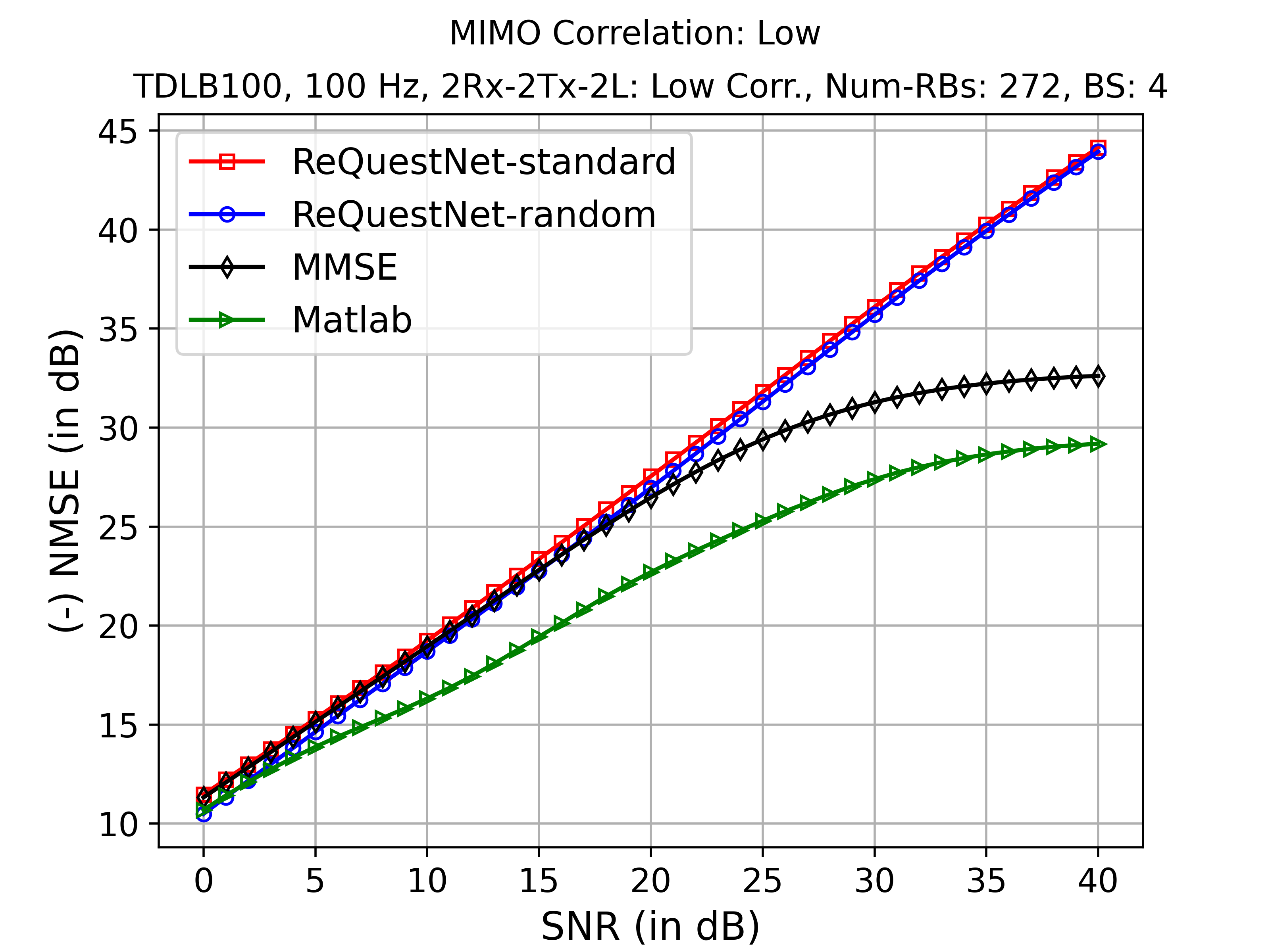}
        \label{fig:low_corr}
    \end{subfigure}
    \begin{subfigure}{0.245\textwidth}
        \centering
        \includegraphics[width=\linewidth]{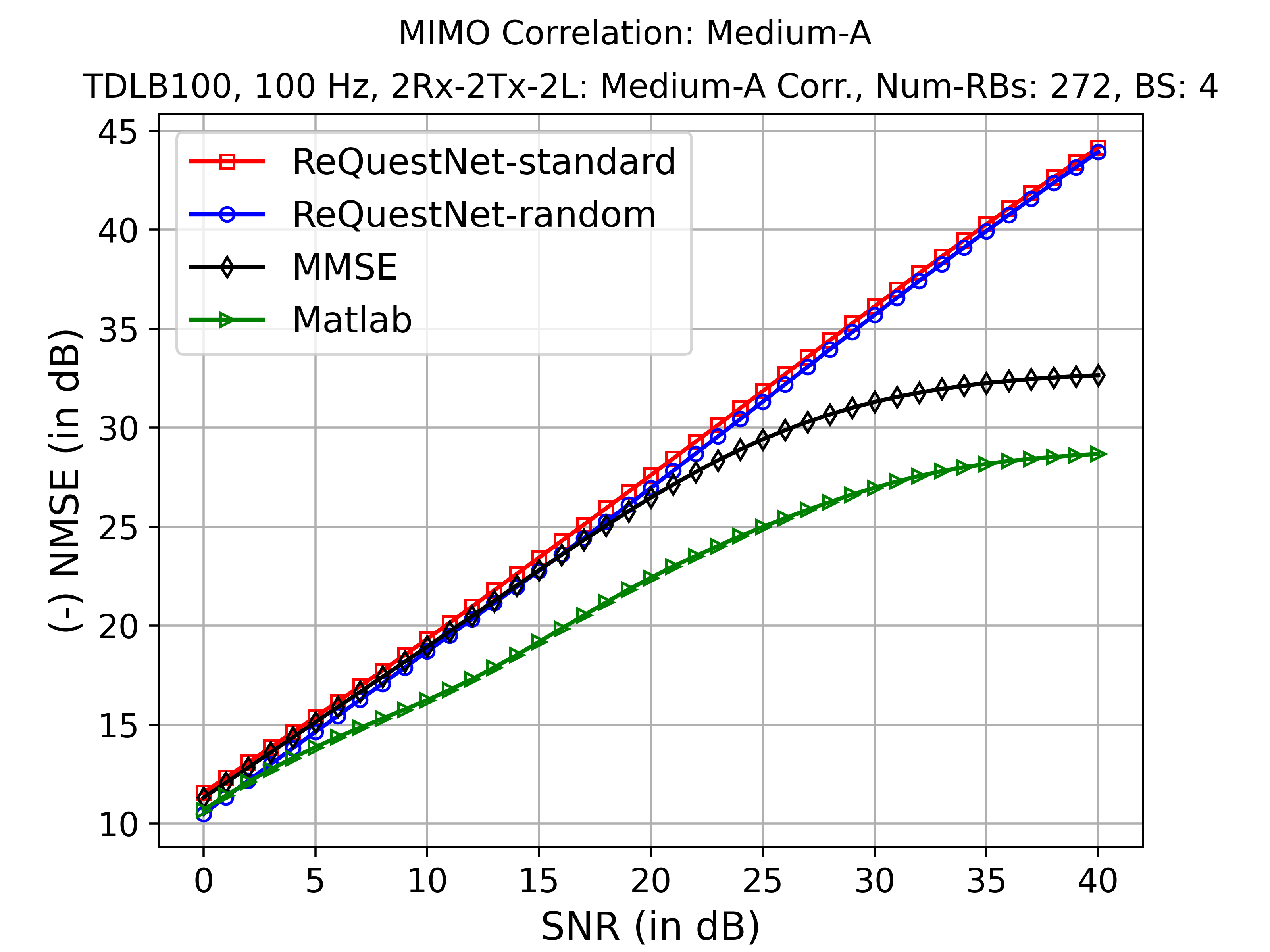}
        \label{fig:med_a_corr}
    \end{subfigure}
    \begin{subfigure}{0.245\textwidth}
        \centering
        \includegraphics[width=\linewidth]{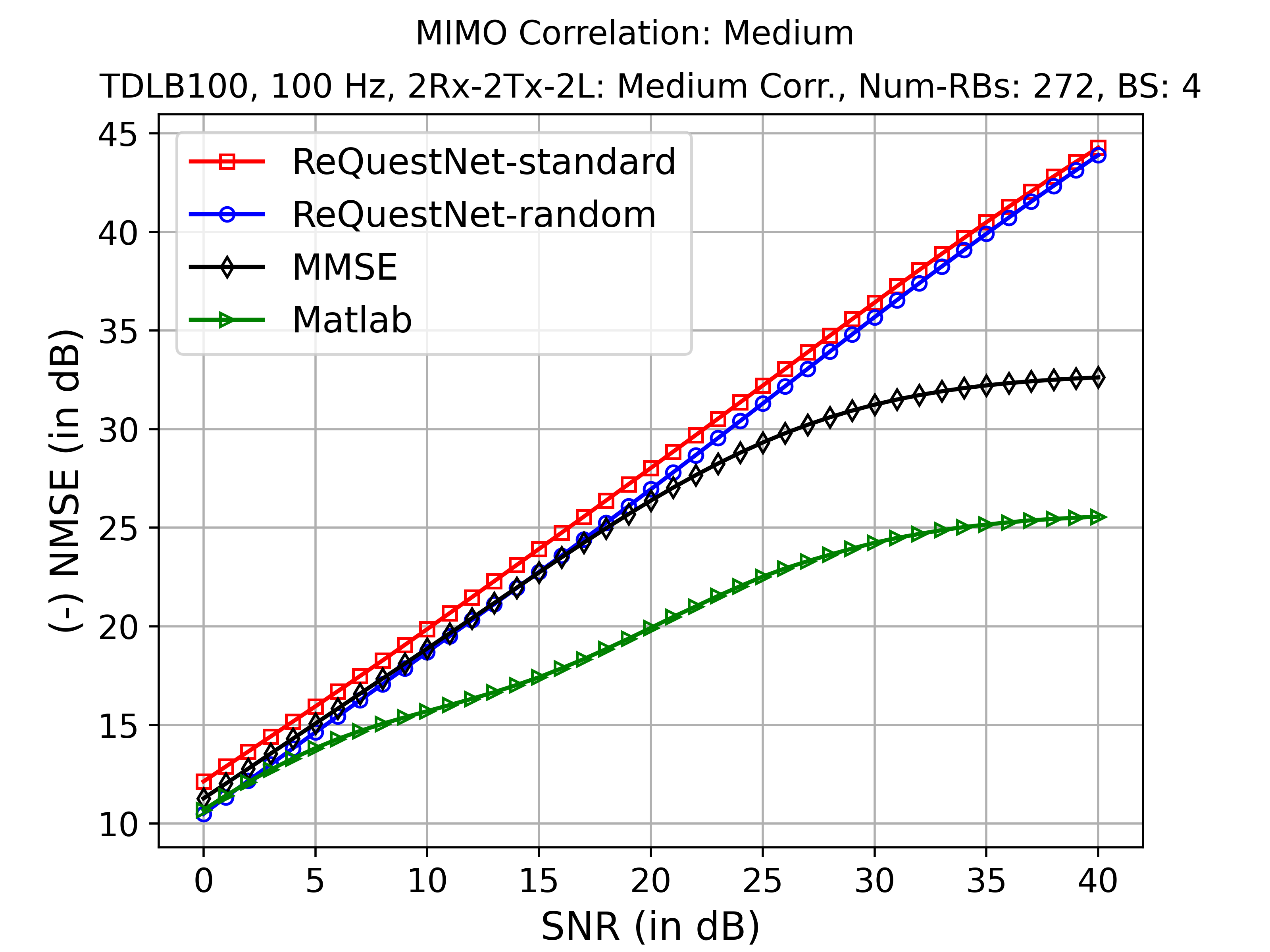}
        \label{fig:med_corr}
    \end{subfigure}
        \begin{subfigure}{0.245\textwidth}
        \centering
        \includegraphics[width=\linewidth]{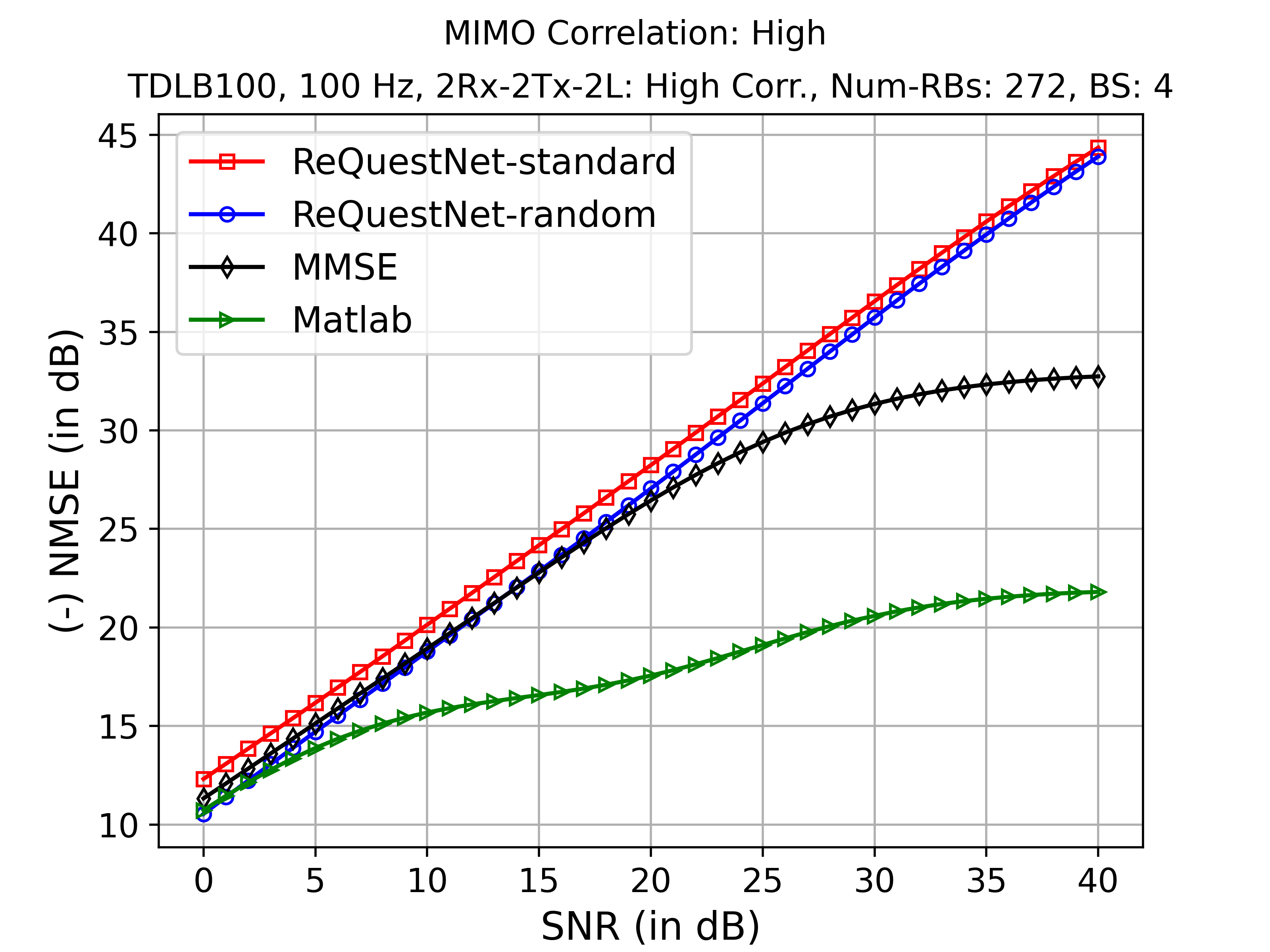}
        \label{fig:high_corr}
    \end{subfigure}
    \caption{($-$NMSE) vs. SNR for different levels of MIMO spatial correlations.}
    \label{fig:mimo_corr_plots}
\end{figure*}

\subsection{Standardization Experiments}
These experiments benchmark \sysname against classical CE methods under standard 3GPP channel configurations.

\begin{table}[!ht]
\centering
\resizebox{\linewidth}{!}
{
\begin{tabular}{|>{\columncolor[gray]{0.8}}c||c|c|c|c|} 
\hline
Category & Profile (TDL) & Doppler (Hz) & Corr & SCS (KHz)\\
\hline
\textbf{Delay} & \textbf{A30}/\textbf{B100}/\textbf{C300} & 100 & Med-A & 30\\ 
\hline
\textbf{Doppler} & TDL-B100 & \textbf{0}/\textbf{30}/\textbf{100}/\textbf{300} & Med-A & 30\\
\hline
\textbf{MIMO corr.} & TDL-B100 & 100 & \textbf{L}/\textbf{MA}/\textbf{M}/\textbf{H} & 30\\
\hline
\textbf{SCS} & TDL-B100 & 100 & Med-A & \textbf{15}/\textbf{30}\\
\hline
\end{tabular}
}
\caption{Configurations for \textit{Standardization} experiments.}
\label{table:standardization_exp_parameters}
\end{table}

\subsubsection{Delay Spread \& Profiles}
We evaluate CE performance across standard 3GPP delay profiles with increasing delay spreads (Row 1, Table.~\ref{table:standardization_exp_parameters}). Shorter delay spreads represent minimal multi-path effects scenarios, while longer delay spreads indicate environments with extensive reflections and scattering affecting CE performance. As shown in Fig.~\ref{fig:delay_profiles_plots}, classical estimators degrade significantly with increasing multipath richness, while \sysname maintains robust performance across all profiles.

\begin{figure}[!ht]
    \centering
    \begin{subfigure}{0.49\linewidth}
        \centering
        \includegraphics[width=\linewidth]{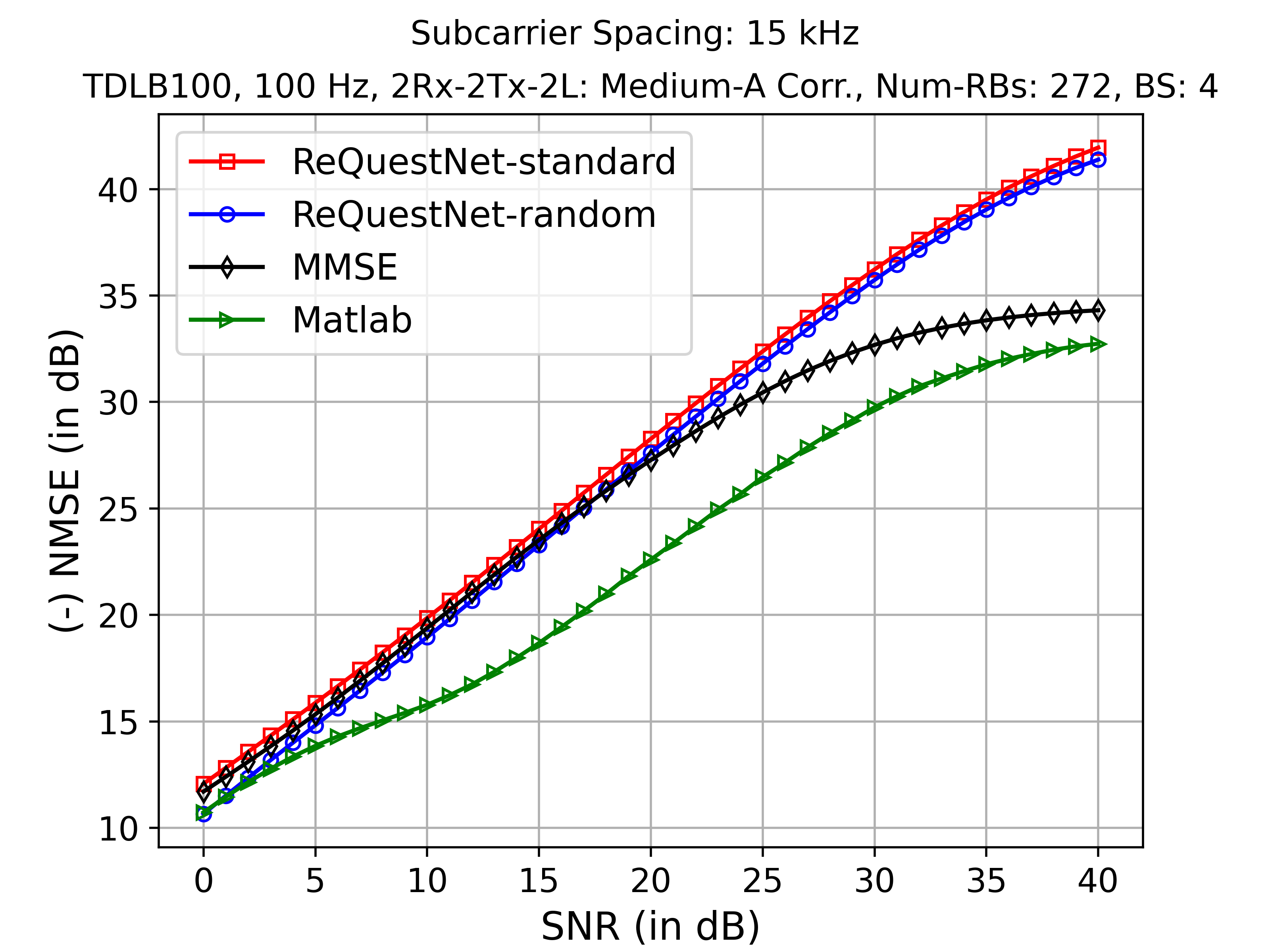}
        \label{fig:scs_15}
    \end{subfigure}
    \begin{subfigure}{0.49\linewidth}
        \centering
        \includegraphics[width=\linewidth]{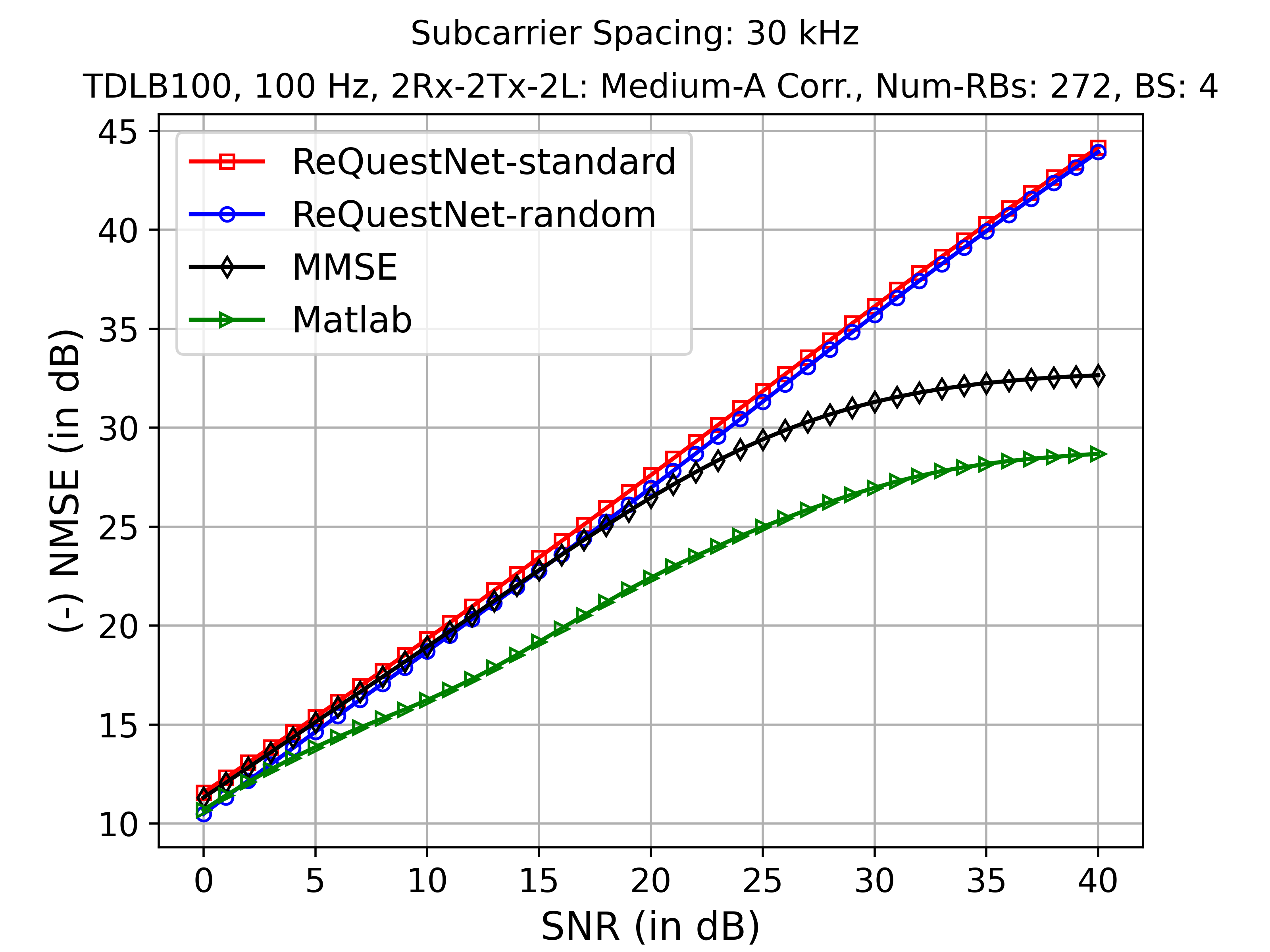}
        \label{fig:scs_30}
    \end{subfigure}
    \caption{($-$NMSE) vs. SNR for different Subcarrier spacing values.}
    \label{fig:scs_plots}
\end{figure}

\subsubsection{Max. Doppler Shift}
Fig.~\ref{fig:doppler_plots} presents results for four Doppler shift values (Row 2, Table.~\ref{table:standardization_exp_parameters}). \sysname consistently outperforms MMSE across the entire Doppler range, showing resilience to mobility-induced channel dynamics.

\subsubsection{MIMO Correlation}
Under increasing spatial correlation (Row 3, Table.~\ref{table:standardization_exp_parameters}), as shown in Fig.~\ref{fig:mimo_corr_plots}, \sysname remains robust even under highly correlated MIMO channels, whereas MMSE performance deteriorates.

\subsubsection{Subcarrier Spacing}
To evaluate compliance with 5G NR numerology, we test \sysname across multiple SCS values (Row 4, Table.~\ref{table:standardization_exp_parameters}). Fig.~\ref{fig:scs_plots} shows that a single trained model generalizes effectively across all SCS values, underscoring its practicality for deployment in heterogeneous network environments.

\subsection{Generalization Experiments}
\begin{figure*}[!ht]
    \centering
    \begin{subfigure}{0.32\textwidth}
        \centering
        \includegraphics[width=\linewidth]{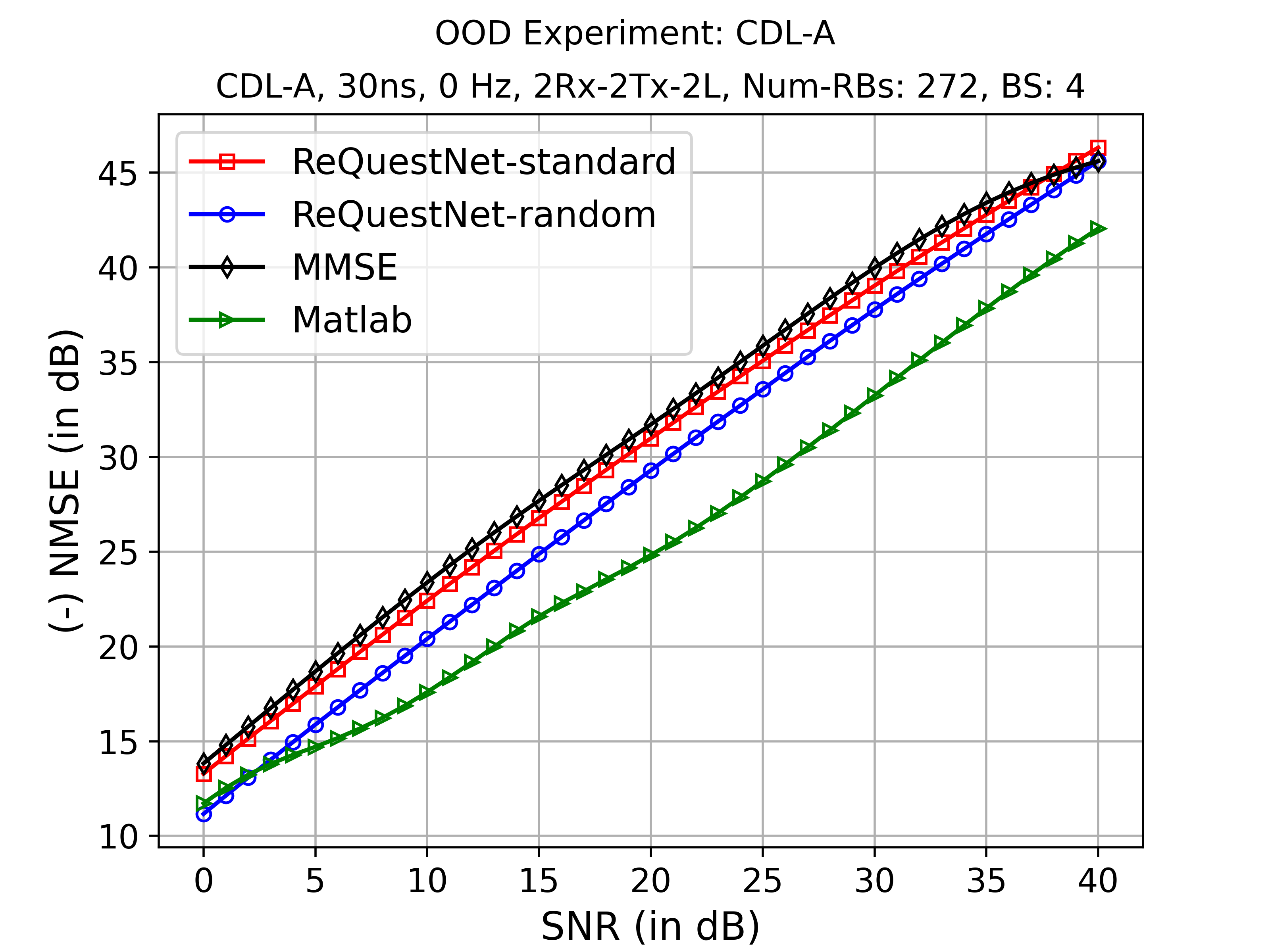}
        \label{fig:cdla_30}
    \end{subfigure}
    \begin{subfigure}{0.32\textwidth}
        \centering
        \includegraphics[width=\linewidth]{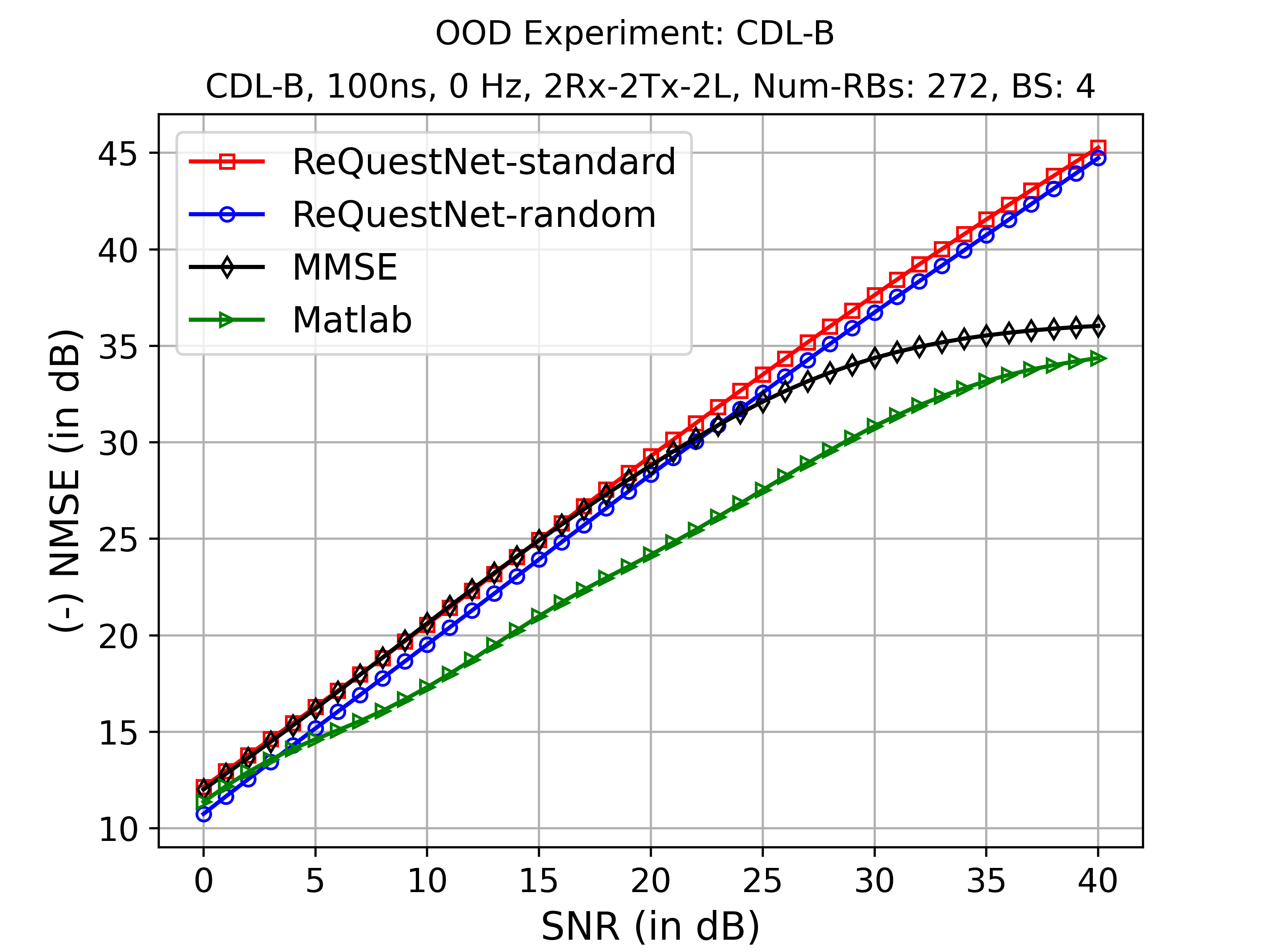}
        \label{fig:cdlb_100}
    \end{subfigure}
    \begin{subfigure}{0.32\textwidth}
        \centering
        \includegraphics[width=\linewidth]{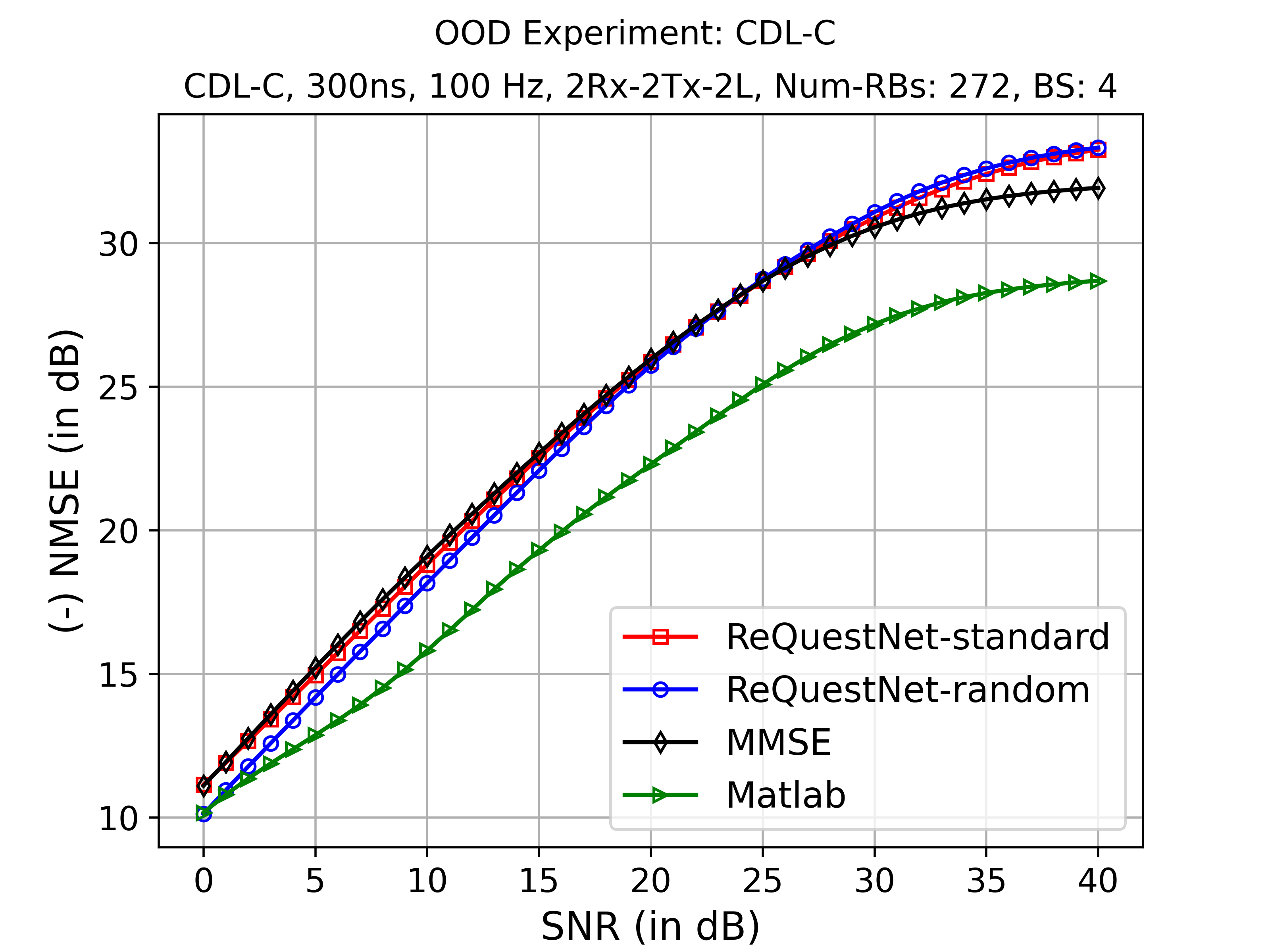}
        \label{fig:cdlc_300}
    \end{subfigure}
    \caption{($-$NMSE) vs. SNR plots illustrating generalizability of ReQuestNet on unseen CDL channel delay profiles.}
    \label{fig:ood_plots}
    \vspace{-5mm}
\end{figure*}

\begin{figure}[!ht]
    \centering
    \begin{subfigure}{0.49\linewidth}
        \centering
        \includegraphics[width=\linewidth]{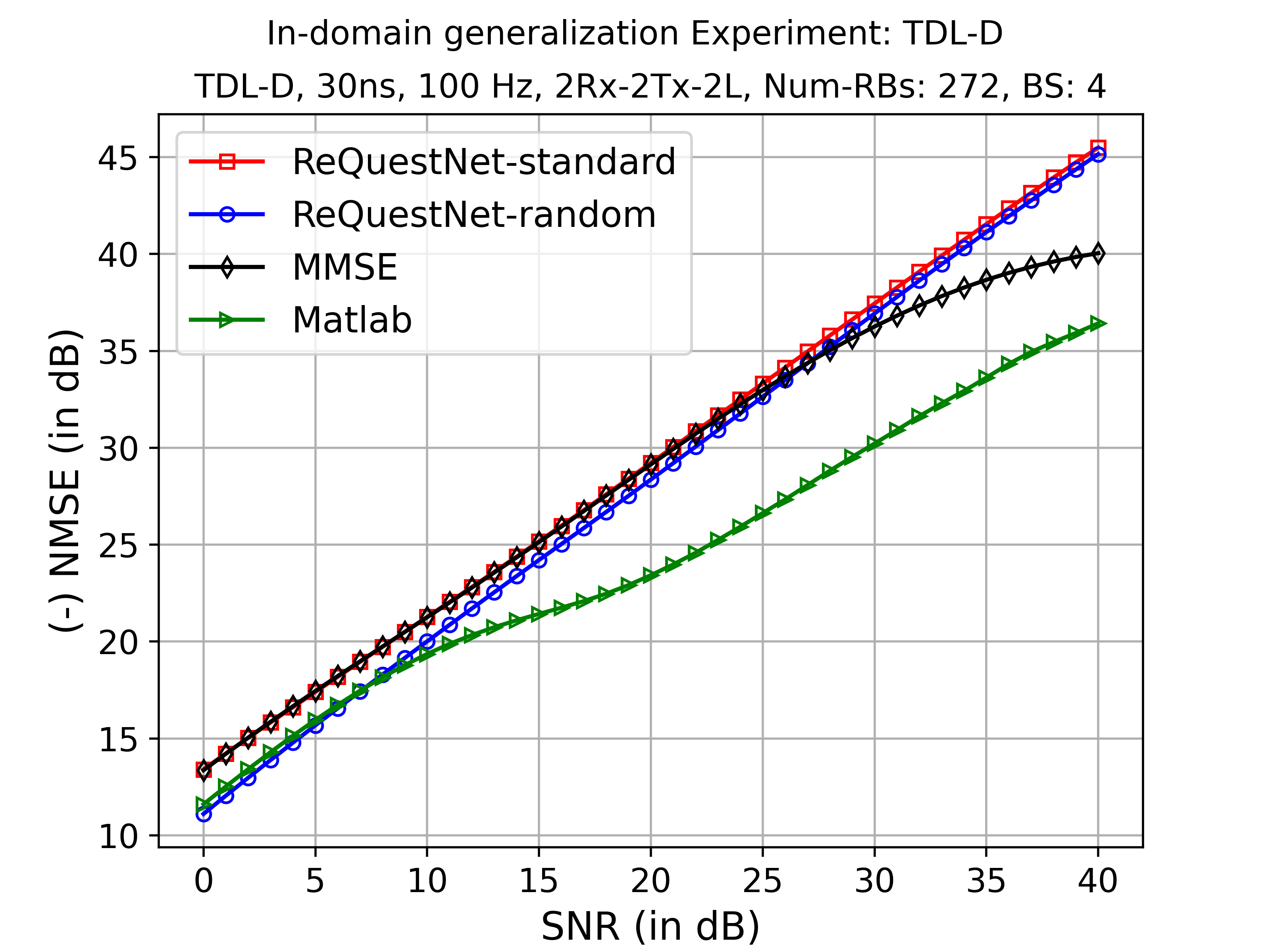}
        \label{fig:tdld_30}
    \end{subfigure}
    \begin{subfigure}{0.49\linewidth}
        \centering
        \includegraphics[width=\linewidth]{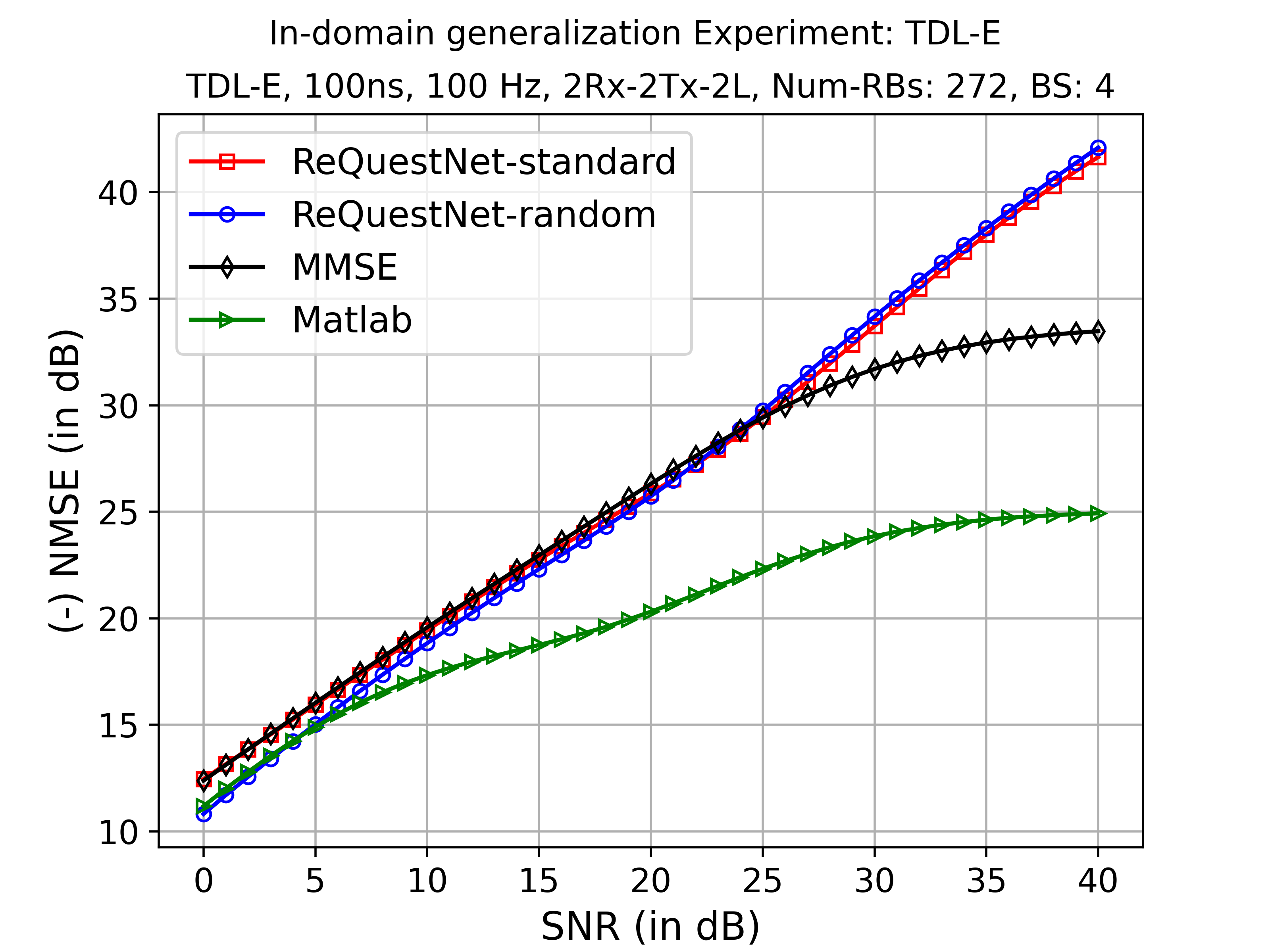}
        \label{fig:tdle_100}
    \end{subfigure}
    \caption{($-$NMSE) vs. SNR plots illustrating generalizability of ReQuestNet on unseen {LOS} \ac{TDL} delay profiles.}
    \label{fig:id_plots}
    \vspace{-5mm}
\end{figure}
The aim here is to examine the generalizability of ReQuestNet when tested on data significantly different from the training regime. The classical MMSE baseline considered here is still the genie MMSE.

\begin{table}[!ht]
\centering
\resizebox{\linewidth}{!}
{
\begin{tabular}{|>{\columncolor[gray]{0.8}}c||c|c|c|c|} 
\hline
Category & Profile (TDL) & Doppler (Hz) & Corr & SCS (KHz)\\
\hline
\textbf{ID} & \textbf{TDL-D/E} & 100 & Med-A & 30\\
\hline
\textbf{OOD} & \textbf{CDL-A/B/C} & 0/0/100 & N/A & 30\\
\hline
\end{tabular}
}
\caption{Configurations for \textit{Generalization} experiments.}
\label{table:generalization_exp_parameters}
\end{table}

This section evaluates the generalization capabilities of \sysname when exposed to channel conditions that differ from those seen during training. While the previous experiments focused on performance under known or closely related configurations, here we assess how well the model extrapolates to novel environments. The genie-aided MMSE estimator remains the baseline for comparison.

We consider two levels of generalization:

\begin{itemize}
    \item \textbf{In-Domain (ID) Generalization:} Evaluates performance on test data that, while not seen during training, remains in the structural vicinity of the training distribution.
    \item \textbf{Out-of-Distribution (OOD) Generalization:} Evaluates extreme generalization performance on test data that is structurally and statistically different from the training distribution.
\end{itemize}

\subsubsection{ID Generalization}
In the ID generalization experiments, we test models trained on non-line-of-sight (NLOS) TDL channel profiles, namely  TDL-A to TDL-C, on line-of-sight (LOS) profiles such as TDL-D and TDL-E. These profiles differ in their delay spread, power delay profile, and multipath richness, but still belong to the same TDL family. As shown in Fig.~\ref{fig:id_plots} and detailed in Row 1 of Table.~\ref{table:generalization_exp_parameters}, \sysname demonstrates strong generalization to these unseen LOS profiles. Despite the structural differences between the training and test profiles, the model maintains high estimation accuracy and continues to outperform the genie-aided MMSE baseline. This indicates that \sysname has learned a robust inductive bias that generalizes well within the TDL family of channels.

\subsubsection{OOD Generalization}
To evaluate more extreme generalization, we test \sysname on clustered delay line (CDL) channel models, which differ significantly from TDL models in terms of delay structure, Doppler characteristics, and spatial correlation. CDL models are more structured and realistic, incorporating angular spreads and cluster-based multipath propagation, making them a challenging testbed for generalization.

These experiments, corresponding to Row 2 of Table.~\ref{table:generalization_exp_parameters} and visualized in Fig.~\ref{fig:ood_plots}, reveal that \sysname remains competitive with the genie-aided MMSE estimator even under such drastic distribution shifts. While a modest performance drop is observed—expected due to the fundamental differences between TDL and CDL models—\sysname still delivers strong performance without any retraining or fine-tuning. This highlights the model’s capacity to generalize beyond its training distribution and adapt to structurally different channel environments. It is important to note that for the \textit{random} variant of \sysname, even the standard TDL profiles used in earlier evaluations are considered OOD. This model is trained exclusively on custom-generated TDL profiles with randomized delay and Doppler characteristics. Despite this, the \textit{random} model performs comparably to the \textit{standard} model across nearly all experiments, including those involving standard 3GPP profiles and CDL channels. This suggests that training on a sufficiently diverse and randomized set of synthetic profiles can endow the model with strong generalization capabilities, even to standardized and real-world channel models.

All generalization experiments were conducted using a single trained model per variant, without any fine-tuning or adaptation to the test conditions. This underscores the versatility and robustness of \sysname as a foundational deep learning model for wireless channel estimation. Its ability to generalize across a wide spectrum of channel conditions—ranging from minor variations to entirely different statistical families—demonstrates its potential for real-world deployment in dynamic and heterogeneous wireless environments. Moreover, this unified modeling approach simplifies deployment pipelines by eliminating the need for model reconfiguration or retraining across different network scenarios. The consistent outperformance of \sysname over the genie-aided MMSE baseline across both ID and OOD settings further reinforces its value as a general-purpose, high-performance solution for channel estimation in 5G and beyond.
\section{Conclusion} \label{sec:conclusions}
We introduced ReQuestNet, a foundational deep learning architecture for channel estimation in 5G NR systems, designed to operate across a wide range of configurations without the need for retraining or manual adaptation. By integrating principles from learned inverse problem solvers, permutation-equivariant modeling, and modular neural design, \sysname effectively addresses the challenges posed by dynamic PRG bundling, varying transmit layers, diverse \ac{DMRS} patterns, and heterogeneous channel conditions. A key innovation lies in its ability to jointly process MIMO spatial streams and differently precoded PRGs, a capability we formally relate to the multi-reference alignment problem. This enables \sysname to unlock processing gains that are inaccessible to classical estimators, including genie-aided MMSE.

Through extensive experiments, we demonstrate that \sysname consistently outperforms traditional baselines across modularity, standardization, and generalization settings. Notably, a single trained model generalizes effectively to both in-domain and out-of-distribution scenarios, including CDL channels not seen during training. These results highlight the potential of \sysname as a unified, high-performance solution for channel estimation. Looking ahead, the architectural principles introduced here—particularly those related to symmetry, modularity, and learned inference—offer a promising foundation for future extensions, including joint channel estimation and demapping, integration with beamforming and scheduling, and adaptation to emerging 6G paradigms. \sysname exemplifies the transformative potential of relatively large AI models in wireless signal processing, bridging theoretical rigor with practical scalability.

\bibliographystyle{IEEEtran}
\bibliography{bibliography}

\end{document}